\documentclass[a4paper, 10pt]{article}
\usepackage[colorlinks=true]{hyperref}
\hypersetup{urlcolor=blue, citecolor=red}
\usepackage{yfonts}
\usepackage[normalem]{ulem}
\usepackage{fancyhdr}
\usepackage{cancel}
\usepackage{latexsym}
\usepackage{color}
\usepackage{amssymb,amsmath,amsfonts,bbm,pifont,upgreek,bbold%,accents
}  
\usepackage[nointlimits]{esint}
\usepackage{graphicx}
\usepackage{tikz}
\usepackage{braket}
\newcommand\beq[1]{\begin{equation}\label{#1} }
\newcommand{\eeq}{\end{equation} }

\renewcommand{\theequation}{\arabic{section}.\arabic{equation}}

\newtheorem{theorem}{Theorem}[section]
\newtheorem{definition}{Definition}[section]
\newtheorem{proposition}{Proposition}[section]
\newtheorem{lemma}{Lemma}[section]

\newtheorem{sublemma}{Sublemma}[section]
\newtheorem{remark}{Remark}[section]
\newtheorem{notationalremark}{Notational Remark}[section]
\newtheorem{corollary}{Corollary}[section]
\newtheorem{assumption}{Assumption}[section]
\newtheorem{claim}{Claim}[section]

\newtheorem{tools}{$\negsp\negsp$}[subsection]

\newcommand\thm[1]{\begin{theorem}\label{#1}}
\newcommand\thmtwo[2]{\begin{theorem}[#1]\label{#2}}
\newcommand\ethm{\end{theorem} }
\newcommand\dfn[1]{\begin{definition}\label{#1} \rm}
\newcommand\dfntwo[2]{\begin{definition}[#1]\label{#2} \rm}
\newcommand\edfn{\end{definition} }
\newcommand\pro[1]{\begin{proposition}\label{#1}}
\newcommand\protwo[2]{\begin{proposition}[#1]\label{#2}}
\newcommand\epro{\end{proposition} }
\newcommand\lem[1]{\begin{lemma}\label{#1}}
\newcommand\lemtwo[2]{\begin{lemma}[#1]\label{#2}}
\newcommand\elem{\end{lemma} }
\newcommand\sublem[1]{\begin{sublemma}\label{#1}}
\newcommand\sublemtwo[2]{\begin{sublemma}[#1]\label{#2}}
\newcommand\esublem{\end{sublemma} }
\newcommand\rem[1]{\begin{remark}\label{#1} \rm}
\newcommand\erem{\end{remark} }
\newcommand\notrem[1]{\begin{notationalremark}\label{#1} \rm}
\newcommand\enotrem{\end{notationalremark} }
\newcommand\cor[1]{\begin{corollary}\label{#1}}
\newcommand\cortwo[2]{\begin{corollary}[#1]\label{#2}}
\newcommand\ecor{\end{corollary} }
\newcommand\asmp[1]{\begin{assumption}\label{#1}}
\newcommand\asmptwo[2]{\begin{assumption}[#1]\label{#2}}
\newcommand\easmp{\end{assumption} }
\newcommand\clm[1]{\begin{claim}\label{#1}}
\newcommand\eclm{\end{claim} }
\expandafter\chardef\csname pre amssym.def
at\endcsname=\the\catcode`\@
\catcode`\@=11
\def\undefine#1{\let#1\undefined}
\def\newsymbol#1#2#3#4#5{\let\next@\relax
 \ifnum#2=\@ne\let\next@\msafam@\else
 \ifnum#2=\tw@\let\next@\msbfam@\fi\fi
 \mathchardef#1="#3\next@#4#5}
\def\mathhexbox@#1#2#3{\relax
 \ifmmode\mathpalette{}{\m@th\mathchar"#1#2#3}%
 \else\leavevmode\hbox{$\m@th\mathchar"#1#2#3$}\fi}
\def\hexnumber@#1{\ifcase#1 0\or 1\or 2\or 3\or 4\or 5\or 6\or 7\or
8\or
 9\or A\or B\or C\or D\or E\or F\fi}
\ifcase\@ptsize
 \font\tenmsb=msbm10
 \font\sevenmsb=msbm7
 \font\fivemsb=msbm5
\or
 \font\tenmsb=msbm10 scaled \magstephalf
 \font\sevenmsb=msbm7 scaled \magstephalf
 \font\fivemsb=msbm5  scaled \magstephalf
\or
 \font\tenmsb=msbm10 scaled \magstep1
 \font\sevenmsb=msbm7 scaled \magstep1
 \font\fivemsb=msbm5 scaled \magstep1
\fi
\newfam\msbfam
\textfont\msbfam=\tenmsb
\scriptfont\msbfam=\sevenmsb
\scriptscriptfont\msbfam=\fivemsb
\edef\msbfam@{\hexnumber@\msbfam}
\def\Bbb#1{\fam\msbfam\relax#1}
\def\widehat#1{\setboxz@h{$\m@th#1$}%
 \ifdim\wdz@>\tw@ em\mathaccent"0\msbfam@5B{#1}%
 \else\mathaccent"0362{#1}\fi}
\def\widetilde#1{\setboxz@h{$\m@th#1$}%
 \ifdim\wdz@>\tw@ em\mathaccent"0\msbfam@5D{#1}%
 \else\mathaccent"0365{#1}\fi}

\def\RIfM@{\relax\ifmmode}
\def\nonmatherr@#1{\errmessage{\string#1\space allowed only in math mode}}
\def\Bbb{\RIfM@\expandafter\Bbb@\else
 \expandafter\nonmatherr@\expandafter\Bbb\fi}
\def\Bbb@#1{{\Bbb@@{#1}}}
\def\Bbb@@#1{\fam\msbfam\relax#1}
\def\setboxz@h{\setbox\z@\hbox}
\def\wdz@{\wd\z@}
\catcode`\@=\csname pre amssym.def at\endcsname
         \usepackage{mathtools}

\DeclareFontEncoding{FMS}{}{}
\DeclareFontSubstitution{FMS}{futm}{m}{n}
\DeclareFontEncoding{FMX}{}{}
\DeclareFontSubstitution{FMX}{futm}{m}{n}
\DeclareSymbolFont{fouriersymbols}{FMS}{futm}{m}{n}
\DeclareSymbolFont{fourierlargesymbols}{FMX}{futm}{m}{n}
\DeclareMathDelimiter{\VERT}{\mathord}{fouriersymbols}{152}{fourierlargesymbols}{147}

\newcommand{\negsp}{\hspace{-.09truecm}}  %%% equivalente a \!

\renewcommand{\P}{{\Pi}   }

\renewcommand{\Im}{{\, \rm Im\, }}

\newcommand{{\cE}}{{\cal  E} }

\newcommand{{\cH}}{{\cal H} }
\newcommand{{\cK}}{{\cal K} }

\newcommand{{\cJ}}{{\cal J}}
\usepackage{color}
\definecolor{yellow}{rgb}{0.99, 0.93, 0.0}

\begin{document}
\title
{\bf Two--Layer Model via Non--Quasi--Periodic Normal Form Theory}

\author{
Gabriella Pinzari$^{1}$, 
Benedetto Scoppola$^{2}$, 
Matteo Veglianti$^{3}$}

%\date{}

\maketitle

\begin{center}
{
\footnotesize
\vspace{0.3cm}$^{1}$  Dipartimento di Matematica “Tullio Levi–Civita”,\\ Università degli Studi di Padova,
\\Via Trieste, 63, 35131 Padova, Italy\\
\texttt{gabriella.pinzari@math.unipd.it}\\

\vspace{0.3cm}$^{2}$  Dipartimento di Matematica,\\
Universit\`a di Roma
``Tor Vergata''\\
Via della Ricerca Scientifica - 00133 Roma, Italy\\
\texttt{scoppola@mat.uniroma2.it}\\

\vspace{0.3cm} $^{3}$ Dipartimento di  Matematica,\\ Universit\`a di Roma
``Tor Vergata''\\
Via della Ricerca Scientifica - 00133 Roma, Italy\\
\texttt{veglianti@mat.uniroma2.it}\\ 
}

\end{center}

\begin{abstract}
The ``two--layer model'' is a $2+\frac{1}{2}$ degrees--of--freedom non--autonomous dynamical system whose lower order expansion exhibits capture in resonance,  numerically detected in a previous paper by the authors~\cite{pinzari2024spin}. In this paper, we reframe the model along the lines of a suitable version of (which we refer to as ``non--quasi--periodic'') normal form theory and provide an explicit amount of the resonance trapping time, which is estimated as exponentially--long, in terms of the small parameters of the system.  
{\bf Key--words:} capture into resonance; non--quasi--periodic normal form theory; friction.

\end{abstract}

\tableofcontents
\renewcommand{\theequation}{\arabic{equation}}
\setcounter{equation}{0}
\section{Introduction}

In a previous paper,~\cite{pinzari2024spin}, we proposed a model (``two--layer model'', in what follows) for the capture into spin-orbit resonance in which the body is composed by two layer, a lighter shell and an heavier core, interacting via a liquid, or viscous, friction. The model is motivated by the study of the capture into 3:2 resonance of Mercury, in which the viscous friction can be related to a melted mantle between a solid crust and a solid kernel, and by the icy Jupiter's satellites, that will be studied in a near future by the JUICE mission~\cite{van2024geophysical}, that are supposed to have a water ocean between the solid icy crust and a rocky core. The model includes a simple -- albeit natural -- description of the different friction felt by the  crust and the core, which is taken  proportional to relative velocity (for a simplified model see~\cite{scoppola2022tides}, other kinds of friction are considered in~\cite{Rochester1970}).
\\
In~\cite{pinzari2024spin} we focused on the study of  the equations of motions of the system (see eq. (25) of that paper) resulting from the lower non-zero terms of a time--averaged  series expansion of the potential in terms of quite natural small parameters of the system: the eccentricity of the orbit, the a-sphericity of the body and the inverse distance from the sun. Notwithstanding the tailored approximations, the motion equations  
which we obtained are still non--trivial, due to nonlinearities.
Quite
surprisingly,
based essentially on numerics, we found that such simplified equations
 provide an account
of a possible mechanism of capture into resonance.  
\\
The purpose of this paper is to understand under which respect 
the neglected higher order terms do not interfere with such description. We remark that, by the occurrence of friction, the model is 
far from being Hamiltonian, whence powerful tools from perturbation theory (like Kolmogorov--Arnold--Moser or Nekhorossev; see below) are not available.
   \\
 Among the recent  theories which deal with friction,  conformally symplectic theory is worth to be mentioned~\cite{wojtkowskiL1998, callejaCL2013, maroS2017}. 
\\
The approach we follow is, in a sense, traditional, in two respects. On one side, as in~\cite{pinzari2024spin}, we start with a
 Lagrangian analysis, as we cannot do differently, 
 due to the occurrence of friction. On the other side, we develop a new analytical tool,  which we refer to as non--quasi--periodic (NQP, hereafter) normal form theory. {Historically, normal form theory has been firstly studied by N. N. Nekhorossev~\cite{nehorosev77, nehorosev79} in connection with the slow motion of action variables whose motion is ruled by a ``close--to--be--integrable'' Hamiltonian  
 \begin{eqnarray*}
    H(I, \varphi)=h(I)+\epsilon f(I, \varphi)\qquad 0<\epsilon\ll 1\,.
    \end{eqnarray*} 
Hamiltonians of this kind are common in the literature (eg, the Hamiltonian of the $n$--body problem, the Euler top, anharmonic interacting oscillators, etc): most of times they are not Liouville--Arnold integrable~\cite{arnold63a}, but are close to systems which are so.
They are widely studied, since the discovery of the so--called Kolmogorov--Arnold--Moser (KAM) theory~\cite{kolmogorov54, moser1962, arnold63c}, which originated a flow of research still not exhausted, which spreads to dissipative and infinite--dimension systems: see, eg,~\cite{wayne1990, kappelerP2003, kuksinM2018, sevryuk2003, callejaCL2017, bertiKM2021, chierchiaP2022, callejaCGL2024} and references therein, for an overview. 
Nekhorossev proved that, along the motions of $H$ the $j^{\rm th}$ action coordinate $I_j$ satisfy an inequality like
  \begin{eqnarray*}
           |I_j(t)-I_j(0)|\le \epsilon^b\qquad {\rm for}\quad |t|\le \frac{1}{\epsilon}\exp{\left(\frac{1}{\epsilon^a}\right)}
           \end{eqnarray*}
           for suitable positive numbers $a$, $b$.
Nekhorossev's papers had a deep impact on the scientific community. After him, many authors  thoroughly studied and progressively clarified the analytic set--up,  both in the original setting~\cite{benettinG1986, poschel93, lochak1992, lochakN1992, bounemouraN2012, guzzoCB2016} or for systems exhibiting an elliptic equilibrium~\cite{fassoGB1998, guzzoFB1998, poschel1999, pinzari13, bounemouraFN2020}, or, finally, for numerical approaches~\cite{cellettiF1996, morbidelliG1997, froescheleGL2000, sansotteraLG2013}.  An extension of normal form theory to non--autonomous Hamiltonian systems with a special decay of the remainder term $f$ has been recently obtained in~\cite{fortunatiW2016}.
Notwithstanding the variety of the recalled analytic results, the occurrence of friction makes them  of no practical use to the two--layer model. Using the  machinery  from~\cite{poschel93}, 
we develop a theory for ODEs equipped with vector--fields where, in the lowest approximation,  part (possibly, none) of the variables has a quasi--periodic motion, while the other part (possibly, all of them) affords dumped oscillations, i.e., oscillations with complex frequencies, whose real part is negative (even though the theory is meaningful for any complex value of the frequency). Previous similar statements  appeared in the unpublished note\footnote{\it arXiv:1710.02689} and, later, in~\cite{ChenP2021}; see~\cite{pinzari23} for a review.

\noindent
Apart from its interest  from a technical point of view, 
we believe that our result is physically meaningful, because   it allows, quite constructively, to ensure that the  motions of the relevant quantities  in the two--layer model are close to such dumped oscillations, for exponentially--long times.  

\noindent
Indeed,  we are able to exhibit an explicit value of the time  $T$ such that for $t<T$ the solution of the linear system stays close, in a suitable norm, to a dumped oscillation, and to compare it with the 
characteristic time $\tau$ ruling the exponential decay, given by the inverse of the modulus of the real part of the frequencies. 
As a matter of fact, if $T>\tau$, the solution will never escape from the equilibrium, at all times.
We remark at this respect that the specificities of the problem at hand  allow us to 
reformulate the underlying, rather complicated, fourth--order eigenvalue equation as second--order ODE 
and
  to treat it via the min--max principle eventually (see  also Remark~\ref{rem: eigenvalues} below). 
\\
This paper is organized as follows. In the next Section~\ref{The Two--Layer Model}, we recall the basic framework of~\cite{pinzari2024spin}, so as to derive the explicit form of the motion equations~\eqref{EL eq}, and state  our main result, Theorem~\ref{main}. In Section~\ref{sec: NFT} we state precisely the aforementioned NQP normal form theory (see Theorems~\ref{iteration lemmaOLD},~\ref{normal form lemma}) and prove Theorem~\ref{main} and, in Sections~\ref{proof of iteration Lemma} and~\ref{proof of NFL} we prove Theorems~\ref{iteration lemmaOLD},~\ref{normal form lemma}, respectively. In Section~\ref{app: Diagonalization}, we provide  the mentioned upper and lower bounds of the size of dumping. Finally, we dedicate Appendix~\ref{Appendix} to recall an abstract result on actions of change of coordinates to vector--fields, which may turn to be useful to non--expert readers.
\vskip.1in
\noindent
{\bf Acknowledgements} We benefited of several comments by A. Celletti, M. Guzzo, C. Lhotka, U. Locatelli, G. Pucacco and A. Sorrentino.\\
BS acknowledges the support of the Italian MIUR Department of Excellence Grant\\ (CUP E83C23000330006).\\
MV has been supported through the ASI Contract n.2023-6-HH.0, Scientific Activities for JUICE, E phase (CUP F83C23000070005).

\section{Lagrangian set--up and result}\label{The Two--Layer Model}
\subsection{The model~\cite{pinzari2024spin}} The two--layer model  is a $2+{\frac{1}{2}}$ degrees--of--freedom dynamical system, constructed as follows. With reference to~\cite[Fig. 1]{pinzari2024spin}, we consider an extended body with total mass $m$ (denoted as $P$, ``planet'', in what follows)  
moving on a plane and undergoing gravity attraction by a 
 point-wise attracting mass $M$  ($S$, ``sun''). For simplicity, we assume that $S$ 
is  fixed in some point of the plane and  that the center of $P$ describes  a Keplerian, elliptic orbit ${\mathcal E}$, with one of its foci at  $S$ and fixed semi--major axis $a$, perihelion direction ${\mathbf i}$  and eccentricity $e$ (we assume the perihelion  is well defined, namely, $e\ne 0$). The position of $P$ on ${\mathcal E}$ is determined by the value of the ``mean anomaly'' $\ell$, which evolves linearly in time, accordingly to Kepler law:
 \begin{eqnarray}\label{ell(t)}
\dot \ell=\omega\qquad \omega:=2\pi \sqrt{\frac{GM}{a^3}}
 \end{eqnarray}
 with $G$ the gravity constant. 
 Concerning the shape and structure of $P$, we assume it  consists of   two  thickless  layers (called ``core''  and  ``shell'' in what follows), both having elliptic shape, but possibly oriented in  different directions. The different orientation of the two ellipses  is physically interpreted as  evidence of mutual friction between the layers, which, as well as in~\cite{pinzari2024spin}, we aim to take into proper consideration. In addition, for the core and the shell we consider only motions which are close to be ``resonant'' (namely, with periods ratios close to a rational number) with  the revolutions of $P$ about $S$. As, due to the friction, the energy is not conserved, the Hamiltonian analysis (and its  powerful machinery) is not an option.
We then proceed with a Lagrangian analysis, as this allows to 
set
 friction forces in  via a Reileigh function $R$. 
To choose the Lagrangian coordinates, we fix a reference frame  with the first axis in the direction ${\mathbf i}$ of the perihelion of ${\mathcal E}$. We denote as $\rho=\rho(a, e, \ell)$  the position of the center of $P$ relatively to $S$; 
 as $\varphi$ and $\nu$ the angles formed by $\rho$ and the semi--major axes directions of the shell and the core.
Since our purpose is to study  
motions for $\varphi$ and $\nu$ which are close to a $2(k/2 +1) : 2(k/2 +1) : 2$ ratio between $\varphi$, $\nu$ and $\ell$ (``spin--orbit resonance''), we introduce the quantities $\gamma$ and $\eta$  via
\begin{eqnarray*}
        \varphi = \frac{k \ell}{2} + \gamma        \text{ \ \ \ and \ \ \ } 
        \nu = \frac{k \ell}{2} + \eta\,.
\end{eqnarray*}
The motion of 
$\gamma$ and $\eta$
 is determined by  the two second--order equations
 \begin{eqnarray}\label{EL eq}
       \left\{
    \begin{array}{lll}
     \displaystyle   \frac{d}{dt} \left( \frac{\partial \mathcal{L}}{\partial \dot\gamma} \right) = \frac{\partial \mathcal{L}}{\partial \gamma} + \frac{\partial R}{\partial \dot\gamma}\\\\
      \displaystyle       \frac{d}{dt} \left( \frac{\partial \mathcal{L}}{\partial \dot\eta} \right) = \frac{\partial \mathcal{L}}{\partial \eta} + \frac{\partial R}{\partial \dot\eta}\,.
   \end{array}
    \right.
\end{eqnarray}
The explicit expressions of ${\cal L}$ and $R$ are
 \begin{eqnarray*} {\mathcal L}= {\mathcal L}_\gamma+ {\mathcal L}_\eta-\widehat{\mathcal V}\,,\qquad 
     R= -\frac{1}{2}\beta \left( \dot{\gamma} - \dot{\eta} \right)^2 - \frac{1}{2}\beta' \left(\dot{\eta} + \frac{k\omega}{2}\right)^2,
 \end{eqnarray*}
with $\beta > 0$ a ``viscous friction coefficient'',  $\beta' > 0$ a ``viscoelastic friction coefficient', and
 \begin{eqnarray}\label{hatVtildeV}
     {\mathcal L}_\gamma      &=& \frac{1}{2} C' \left[ \frac{k}{2}\omega + \dot{\gamma} + \dot\vartheta(\ell, \omega, e)\right]^2 +  \frac{3}{8}\omega^2(B'-A') a_k(e)e^k\cos 2\gamma-\widetilde{\mathcal V}'(\ell, \gamma, e)\nonumber\\
     {\mathcal L}_{\eta} &=& \frac{1}{2} C \left[ \frac{k}{2}\omega + \dot{\gamma} + \dot\vartheta(\ell, \omega, e)\right]^2 +  \frac{3}{8}\omega^2(B-A) a_k(e)e^k\cos 2\eta-\widetilde{\mathcal V}(\ell, \eta, e) \nonumber\\
 \widetilde{\mathcal V}\,,&& \widetilde{\mathcal V}'={\rm O}\left(\frac{r^2}{a^3}\right)\,,\quad     \widehat{\mathcal V}={\rm O}\left(\frac{r^3}{a^4}\right)
\end{eqnarray}
where
 $\widetilde{\mathcal V}(\ell, \eta, e)$, $\widetilde{\mathcal V}'(\ell, \eta, e)$ have vanishing $\ell$--average
and with   $r$  being the average radius of $P$. 
To lighten notations, we
introduce the homogeneous quantities
\begin{eqnarray*}c_1&:=&\frac{3}{4} \frac{B'-A' }{C'}\frac{\omega^2} e^k a_k(e)\,,\quad \theta:=\frac{\beta}{C'}\nonumber\\
c_2&:=&\frac{3}{4} \frac{B-A }{C}\omega^2 e^k a_k(e)\,,\quad \epsilon:=\frac{\beta}{C}\,,\quad \upupsilon:=\frac{\beta'}{C}\,,\quad -v_0:=\frac{k}{2}\omega_0\nonumber\\
\widetilde P_\gamma&:=&-\frac{\partial_\gamma\widetilde{\cal V}'}{C'}\,,\quad \widehat P_\gamma:=-\frac{\partial_\gamma\widehat{\cal V}'}{C'}\,,\quad
\widetilde P_\eta:=-\frac{\partial_\eta\widetilde{\cal V}}{C}\,,\quad \widehat P_\eta:=-\frac{\partial_\eta\widehat{\cal V}}{C}
\end{eqnarray*}
The coefficients $c_1$, $c_2$ are related to the geometry and the order of resonance, while $\theta$, $\epsilon$ and $\upupsilon$ are related to the friction coefficients $\beta$, $\beta'$. From the physical meaning of such coefficients, in this paper we shall always regard
\begin{eqnarray}\label{friction coefficients}
\theta>\epsilon>\upupsilon
\end{eqnarray}
even though more precise quantitative relations will be specified.\\
With the above definitions and notations,  we rewrite Equations~\eqref{ell(t)} and~\eqref{EL eq} in the form of first order ODEs system
\begin{eqnarray}
    \label{eq_2equation_coupledNEW}
\left\{\begin{array}{llll}
              \displaystyle  \dot\gamma= p_\gamma\\
       \displaystyle            \dot p_\gamma = - c_1  \sin{2 \gamma} - \theta (p_\gamma - p_\eta)+\widetilde P_\gamma+\widehat P_\gamma\\
                 \displaystyle  \dot\eta=p_\eta\\
        \displaystyle            \dot p_\eta = - c_2 \sin{2\eta} + \epsilon (p_\gamma - p_\eta) -\upupsilon (p_\eta - v_0)+\widetilde P_\eta+\widehat P_\eta\\
        \dot\ell=\omega
    \end{array}\right.
\end{eqnarray} 
Neglecting the $P$'s releases\footnote{Compare Eq. (25) in~\cite{pinzari2024spin} and the comment below.} the system considered in~\cite{pinzari2024spin}.
The system~\eqref{eq_2equation_coupledNEW} will be referred to as {\it full system} in what follows. Such locution is to be understood as opposite to {\it linearized system}, which is a 
further simplification, not considered in~\cite{pinzari2024spin}, and now we describe.
\subsection{The linearized system and result}
We consider the point 
 $(0, 0, \eta_0, 0)$, where
\begin{eqnarray*} \eta_0:=\frac{1}{2}\sin^{-1}\left(\frac{\upupsilon}{c_2}v_0\right)\quad \mod \pi\,.\end{eqnarray*}
The point $(0, 0, \eta_0, 0)$ is an equilibrium of the vector--field obtained from the first four equations in~\eqref{eq_2equation_coupledNEW} by neglecting all the $P$'s. It 
is
 well--defined provided that \begin{eqnarray*}\frac{\upupsilon}{c_2}|v_0|
<1\,.\end{eqnarray*}
An expansion about such equilibrium leads to the system
\begin{eqnarray}\label{linear system}
\left\{\begin{array}{llll}
              \displaystyle  \dot\gamma=p_\gamma\\
       \displaystyle            \dot p_\gamma = - 2c_1\, \gamma - \theta p_\gamma +\theta p_{\psi}
       +\breve P_\gamma+\widetilde P_\gamma+\widehat P_\gamma
       \\
                 \displaystyle  \dot \psi=p_{\psi}\\
        \displaystyle            \dot p_{\psi} = - 2\bar c_2 \,\psi + \epsilon p_\gamma  -\delta p_{\psi} +\breve P_u+\widetilde P_u+\widehat P_u\\
        \dot\ell=\omega
    \end{array}\right.
\end{eqnarray}
where $\eta$ has been changed with $\psi:=\eta-\eta_0$, \begin{eqnarray*}\delta:=\epsilon+\upupsilon\,,\ \bar c_2:=c_2\cos2\eta_0
\end{eqnarray*}
$\widetilde P_\gamma$, $\widehat P_\gamma$, $\widetilde P_u$, $\widehat P_u$ denoting (with abuse) the previous functions in the new variables, and
\begin{eqnarray}\label{system}
\breve P_\gamma(\gamma, \psi,  \ell):=- c_1 \sin2\gamma+2c_1\gamma\,,\quad \breve P_u(\gamma, \psi,  \ell):=
         - c_2\sin{2(\psi+\eta_0)}+
        \upupsilon v_0+
        2\bar c_2\,\psi
        \end{eqnarray}
        being the higher order terms released from the expansion. 
        From now on, we shall neglect to write
    the ``bar'' in~\eqref{linear system}. Neglecting all the $P$'s  the system decouples as a linear one involving the ``slow'' variables $(\gamma, p_\gamma, \psi, p_{\psi})$, and hence named {\it linearized system},
    \begin{eqnarray}\label{linear system1}
\left\{\begin{array}{llll}
              \displaystyle  \dot\gamma=p_\gamma\\
       \displaystyle            \dot p_\gamma = - 2c_1\, \gamma - \theta p_\gamma +\theta p_{\psi}
       \\
                 \displaystyle  \dot \psi=p_{\psi}\\
        \displaystyle            \dot p_{\psi} = - 2c_2 \,\psi + \epsilon p_\gamma  -\delta p_{\psi}    \end{array}\right.
\end{eqnarray}
plus the  equation~\eqref{ell(t)} for the ``fast'' variable $\ell$, left apart. We denote as $L$ the matrix of the coefficients of~\eqref{linear system1}. We have 
\begin{proposition}\label{prop: asymptotically stable} 
Under an open and generic condition (i.e., if  the resolvent of the characteristic polynomial of $L$ does not vanish), if the inequality in~\eqref{friction coefficients} and 
\begin{eqnarray}\label{negativity condition}
\epsilon\ne0\,,\quad \theta^2<\frac{8}{9}\min\{c_1\,,c_2\}
\end{eqnarray}
hold, then $L$ admits two distinct complex--conjugated couples of eigenvalues $\lambda_j$, with strictly negative real part and non--vanishing imaginary part. More precisely, the following bound holds
\begin{eqnarray}\label{bounds lambdai}
{\, \rm Re\, }\lambda_j\subset\left[-\frac{3}{2}\theta\,,-\frac{\upupsilon}{3}\right]\,.
\end{eqnarray}
\end{proposition}
\noindent
The  proof of Proposition~\ref{prop: asymptotically stable} is provided in 
 Section~\ref{app: Diagonalization}.
 \vskip.1in
 \noindent
Proposition~\ref{prop: asymptotically stable}   implies that the motions of $\gamma$ and $\psi$ along the solutions   of the linearized system  are given by 
\begin{eqnarray}\label{damped}
\gamma_*(t):=\sum_{j=1}^4b_{1j} e^{\lambda_j t}\,,\qquad \psi_*(t):=\sum_{j=1}^4b_{3j} e^{\lambda_j t} 
\end{eqnarray}
where $b=(b_{ij})$ is such that $b^{-1}L b$ is in diagonal form. 

\vskip.1in
\noindent
We define
\begin{eqnarray*}
\mu_0&:=&{\, \rm const}\max\left\{c_1\,,\  c_2\,,\ 
\varepsilon_0\,,\ \epsilon\varepsilon_0\,,\ \delta\varepsilon_0\,,\ |v_0|\upupsilon\,,\ \frac{r^2}{a^3\varepsilon_0\min\{C\,,\ C'\}}
\right\}\nonumber\\
\mu_1&:=&{\, \rm const}\max\left\{\frac{\mu_0^2}{\omega}\,,\ \mu_0\frac{r}{a}\,,\ c_1\varepsilon_0^3\,,\ c_2\varepsilon_0^2\right\}\nonumber\\
\gamma_1&:=&\min\left\{\frac{\upupsilon}{3}\,,\ \omega\right\}\nonumber\\
T&:=&{\, \rm const}^{-1}\frac{\varepsilon}{\varepsilon_*\mu_1}\,
 e^{\frac{\gamma_1}{\mu_1}}
\end{eqnarray*}
where the r\^ole of ``${\, \rm const}$'', $\varepsilon$, $\varepsilon_*$
is specified as follows:

\begin{theorem}\label{main}
Under the generic assumptions of Proposition~\ref{prop: asymptotically stable}, there exists a value of ``${\, \rm const}$'' and a positive number ${\varepsilon_*}$ of $(0,0,\eta_0, 0)$ such that, for all $\varepsilon>0$ such that\begin{eqnarray}\label{omega0}
\frac{\mu_0}{\omega}\le\varepsilon\,,\ \frac{\mu_1}{\gamma_1}\le\varepsilon\end{eqnarray}
  all the solutions of the system of ODEs~\eqref{linear system} with initial datum in $B_{\varepsilon_*}$ 
verify
\begin{eqnarray}\label{asymptotically stable}
|\gamma(t)-\hat\gamma(t)| <\varepsilon\,,\quad |\psi(t)-\hat \psi(t)|<\varepsilon\qquad \forall\ |t|< T
\end{eqnarray}
where $\hat\gamma(t)$, $\hat \psi(t)$ have the expression in~\eqref{damped}, with $\lambda_j$ replaced by suitable $\hat\lambda_j$ verifying
\begin{eqnarray*}
{\, \rm Re\, }\hat\lambda_j<0\,,\quad|\hat\lambda_j-\lambda_j|\le \frac{\mu_1}{2}\,.
\end{eqnarray*}
\end{theorem}

\section{Proof of Theorem~\ref{main} via NQP Normal Form Theory}\label{sec: NFT}
The proof of Theorem~\ref{main} uses  a formulation of normal form theory for vector--fields, carefully designed around the system~\eqref{linear system}. More precisely, it is based on two results (Theorem~\ref{iteration lemmaOLD} and Theorem~\ref{normal form lemma} below) which  here we quote together with the necessary background of notations and definitions. 
While the proof of such results is deferred to the next sections, here we prove how Theorem~\ref{main} follows from them.

\vskip.1in
\noindent 
We first fix some notation and definition. The former defines the functional setting and suitable norms, so--called {\it weighted}.
\begin{definition}\rm
\item[\tiny\textbullet] For a given set $A\subset {\mathbb R}^p$ and $r>0$, we let
\begin{eqnarray*}
A_r:=\bigcup_{x\in A}B_r(x)
\end{eqnarray*}
where $B_r(x)$ is the complex ball with radius $r$ centered at $x$:
\begin{eqnarray*}
B_r(x):=\big\{z\in {\mathbb C}^p:\ |z-x|<r\big\}
\end{eqnarray*}
Here, $|\cdot|$ is some fixed norm of $\mathbb C^p$.
\item[\tiny\textbullet] We denote as ${\mathcal O}_{u}$, with $u:=(\varepsilon, s)$,  the space of vector--fields \begin{eqnarray}\label{Z}Z=(Z_1\,,\ldots\,,Z_{m+n}):\quad V_u:=B^m_\varepsilon\times {\mathbb T}^n_s\to{ \mathbb C}^{m+n}
\end{eqnarray}
which are holomorphic  on $V_{u_0}$, $u_0=(\varepsilon_0, s_0)$, with some $\varepsilon_0>\varepsilon$, $s_0>s$.
\item[\tiny\textbullet]  If 
\begin{eqnarray}\label{vec expansion}Z_h=\sum_{\alpha,  k} z_{\alpha k}^h\zeta^\alpha e^{{\rm i}(k\cdot \varphi)}\end{eqnarray}
denotes the Taylor--Fourier expansion, we define
the {\it weighted norms} as
\begin{eqnarray}\label{normXOLD}|X|_{u}^{w}&:=&\sum_{h=1}^{m+n} w^{-1}_h|Z_h|_{u}\,,\quad 
\VERT X \VERT_{u}^{w}:=\sum_{h=1}^{m+n} w^{-1}_h\|Z_h\|_{u}\end{eqnarray}
 where
\begin{eqnarray*}|Z_h|_{u}:=\sup_{V_u}|Z_h|\,,\quad
\|Z_h\|_{u}:= \sum_{\alpha,  k} |z_{\alpha k}^h|\varepsilon^{|\alpha|_1} e^{|k|_1 s}\end{eqnarray*}
and with
$w=(w_1$, $\ldots$, $w_{m+n})\in {\mathbb R}_+^{m+n}$  the {\it weights}.
\end{definition}
We next define $(\gamma, {{\Lambda}}, K)$--nonresonance; $T_K$ and $\Pi_{{\Lambda}}$ projectors.
\begin{definition} 
\label{nonresonance}\rm
\item[\tiny\textbullet]  Fix $\gamma>0$  and ${{\Lambda}}\subset 
{\mathbb N}^m\times  {\mathbb Z}^n$, with $0\in{{\Lambda}}$.  We say that  $(\lambda, \omega)$ is $(\gamma, {{\Lambda}}, K)$--{\it nonresonant} if  $$|\lambda\cdot \alpha+{\rm i}\omega\cdot k|\ge \gamma\qquad \forall\ 
(\alpha, k)\notin {{\Lambda}}\,,\quad |(\alpha, k)|_1\le K\,.$$
\item[\tiny\textbullet] For each $1\le h\le n+m$, let $p_h=(p_{h1}, \ldots, p_{h, {n+m}})\in  {\mathbb N}^m\times  {\mathbb Z}^n$ be defined so that
\begin{eqnarray}\label{ph}p_{hj}=\left\{\begin{array}{llll}
\delta_{hj}\quad &{\rm if}\ 1\le h\le m\\
0&{\rm if}\ m+1\le h\le m+n
\end{array}\right.\end{eqnarray}
where $\delta_{hj}$ is the Kronecker symbol. Given ${{\Lambda}}\subset {\mathbb N}^m\times  {\mathbb Z}^n$, with $0\in{{\Lambda}}$ and $h=1\,,\ldots\,,m+n$, let ${{\Lambda}}_h\subset {\mathbb N}^m\times  {\mathbb Z}^n$ be the {\it $p_h$--translated lattice}
$${{\Lambda}}_h:={{\Lambda}}+p_h\,.$$
\item[\tiny\textbullet] Given $Z$ as in~\eqref{Z}--\eqref{vec expansion}, the projectors $T_K Z$, $\Pi_{{\Lambda}}Z$ will denote the vector--fields defined via
\begin{eqnarray*}
(T_KZ)_h:=\sum_{|(\alpha,  k)|\le K} z_{\alpha k}^h\zeta^\alpha e^{{\rm i}(k\cdot \varphi)}\qquad (\Pi_{{\Lambda}}Z)_h:=\Pi_{{\Lambda}_h}Z_h:=\sum_{(\alpha,  k)\in{{\Lambda}}_h} z_{\alpha k}^h\zeta^\alpha e^{{\rm i}(k\cdot \varphi)}\,.
\end{eqnarray*}
 \end{definition}
 \vskip.1in
 \noindent
 We now quote two results (Theorem~\ref{iteration lemmaOLD} and~\ref{normal form lemma}) concerning a system of ODEs \begin{eqnarray}\label{oldVF}
\dot x=X(x)
\end{eqnarray}
where 
$x:=(\zeta, \varphi)\in B\times {\mathbb T}^n$, with $B\subset {\mathbb R}^m$ is a neighborhood of $0=(0, \ldots, 0)$,
$X(x)$ is a vector--field   having the form
\begin{eqnarray}\label{X}X(x)=N(x)+ P(x)\end{eqnarray}
where $N(x)$ is $\varphi$--independent
and given by
\begin{eqnarray}\label{N}
N(\zeta)=\left(
\begin{array}{ccc}
\lambda_1 \zeta_1\\
\vdots\\
\lambda_m\zeta_m\\
\omega_1\\
\vdots\\
\omega_n
\end{array}\right)\end{eqnarray}
with suitable $\lambda\in {\mathbb C}^m$, $\omega\in {\mathbb C}^n$.
Then we have
\begin{theorem}\label{iteration lemmaOLD}
Let $1\le K\in \mathbb N$, $X\in {\mathcal O}_{u}$ be as in~\eqref{X}, with $N$ as in~\eqref{N},  $u=(\varepsilon, s)$, $w=(\rho,\sigma)<u/2$, $0\in{\Lambda}\subset {\mathbb Z}^{m+n}$. Assume that $\upomega=(\lambda, {\rm i}\omega)$ is $(\gamma, {{\Lambda}}, 2K)$--nonresonant and that $P$  is so small that
\begin{eqnarray}\label{smallcond}
e \gamma^{-1} \VERT P\VERT^{w} _{u}<1\end{eqnarray}
 Then there exists a holomorphic change of coordinates
 \begin{eqnarray*}
 \phi_+:x_+=(\zeta_+, \varphi_+)\to x=(\zeta, \varphi)
 \end{eqnarray*}
 which carries $X$ to $X_+\in{\mathcal O}_{u-2w}$, with  
$X_+=N+G_++
P_+$, where \begin{eqnarray}\label{G+OLD}G_+=\Pi_{{\Lambda}} T_K P\,.\end{eqnarray} Moreover, there exists $Y\in {\mathcal O}_{u}$  such that
$X_+:=e^{{\mathcal L}_Y}X$ and
\begin{eqnarray}\label{P+OLD}
 \VERT P_+\VERT^{w}_{u-2w}
\le \frac{1
}{1-e \gamma^{-1}
\left\VERT P\right\VERT^w_{u}}\left(
e \gamma^{-1}\VERT P\VERT^w_{u}\VERT   P\VERT^w_{u-w}+e^{-K \tau}\left\VERT P\right\VERT^w_{u}\right)
\end{eqnarray}
with $\tau$ as in~\eqref{tau} and $\VERT Y\VERT^w_{u}\le \gamma^{-1}\VERT P\VERT^w_{u}$. Finally, the transformation $\phi_+$ verifies
\begin{eqnarray}\label{close to id OLD}
|\phi_+-{\rm id}|_{u-2w}^w\le \gamma^{-1}\VERT P\VERT_u^w
\end{eqnarray}
\end{theorem}

\begin{theorem}\label{normal form lemma} There exists $C_*>0$ such that the following holds.
Let $X=N
+P\in {\mathcal O}_{u}$, with $u=(\varepsilon, s)$, $N$ as in~\eqref{N}, $w=(\rho,\sigma)<u/4$, $0\in{\Lambda}\subset{\mathbb Z}^{m+n}$. Put $\bar\sigma:=\min\{\sigma, \rho/\varepsilon\}$.
Assume that $\upomega=(\lambda, {\rm i}\omega)$ is $(\gamma, {{\Lambda}}, 2K)$--non resonant, with
\begin{eqnarray}\label{Ksigma large} K\bar\sigma\ge \log (12)\end{eqnarray}
 and that $P$  is so small that
\begin{eqnarray}\label{smallcondNEWNEW}
C_* K \bar\sigma \gamma^{-1} \VERT P\VERT^{w} _{u}<1\qquad \end{eqnarray}
 Then there exists a holomorphic transformation of coordinates \begin{eqnarray*}\phi_*:\quad V_{u-4w}\to V_{u}\end{eqnarray*}  which carries $X$ to $$X_*=N+G_*+
P_*\in {\mathcal O}_{u-4w}$$ with $G_*$ verifying 
$G_*=\P_{{\Lambda}}T_K G_*$
and \begin{eqnarray*}
\VERT G_*-\Pi_{{\Lambda}}T_K P\VERT^w_{u-4w}\le
8
e \gamma^{-1}\left(\VERT P\VERT^w_{u}\right)^2\end{eqnarray*}
and
 $P_*$ ``small'':
\begin{eqnarray*} \VERT P_*\VERT^{w}_{u-4w}
\le
e^{-K \bar\sigma/4}\left\VERT P\right\VERT^w_{u}
\end{eqnarray*}
Finally, the transformation $\phi_*$ verifies
\begin{eqnarray}\label{phi* close to id}
|\phi_*-{\rm id}|_{u-4w}^w\le 2\gamma^{-1}\VERT P\VERT _{u}^{w}\,.\end{eqnarray}

\end{theorem}
\noindent
We can now provide the

\par\medskip\noindent{\bf Proof\ } {\bf of Theorem~\ref{main}}
We let $\zeta_0:=(\gamma, p_\gamma, \psi, p_\psi)$, $\varphi_0:=\ell$, $x_0:=(\zeta_0, \varphi_0)$. \begin{eqnarray}\label{A}N_0=\left(\begin{array}{cc}0\\\omega
\end{array}\right)\,,\quad P_0:=\left(\begin{array}{cc}L\zeta_0\\0
\end{array}\right)+\breve P+\widetilde P+\widehat P\end{eqnarray}
 where the matrix $L$ as well as the components of $\breve P$, $\widetilde P$ and $\widehat P$ are defined via the right hand side of~\eqref{linear system}. Then the vector--field at right hand side of~\eqref{linear system} is
\begin{eqnarray}\label{system0}
X_0(x_0)=N_0+P_0(x_0)
\end{eqnarray}
We next proceed in four steps. In steps 2 and 3, 
``${\, \rm const}$'' will be a suitably large number, independent of  $\omega$,  $\theta$, $\epsilon$, $\upupsilon$.

\paragraph*{Step 1: application of Theorem~\ref{iteration lemmaOLD}}  
We fix $u=u_0=(\varepsilon_0, s_0)$ so that $P_0$ is real--analytic
in the domain
\begin{eqnarray*}
\zeta_0\in B^4_{\varepsilon_0}\,,\ \varphi_0\in {\mathbb T}_{s_0}
\end{eqnarray*} 
 and choose the weights $w_0=\frac{u_0}{4}$. We
  can bound (see equations~\eqref{eq_2equation_coupledNEW} and~\eqref{hatVtildeV})
\begin{eqnarray}\label{C}\VERT P_0\VERT _{u_0}^{w_0}\le {\, \rm const}\max\left\{c_1\,,\  c_2\,,\ 
\varepsilon_0\,,\ \epsilon\varepsilon_0\,,\ \delta\varepsilon_0\,,\ |v_0|\upupsilon\,,\ \frac{r^2}{a^3\varepsilon_0\min\{C\,,\ C'\}}
\right\}=:\mu_0\,.
\end{eqnarray}
 We aim to apply Theorem~\ref{iteration lemmaOLD} to $N=N_0$, $P=P_0$ as in~\eqref{system0}. Condition~\eqref{smallcond} is satisfied, due to~\eqref{omega0}.
The frequency $\upomega_0=(0, {\rm i}\omega)$ is $(\gamma_0, {{\Lambda}}_0, 2K_0)$--non resonant with \begin{eqnarray}\label{gamma0}\gamma_0=\omega\end{eqnarray} for all $K_0\in{ \mathbb N}$. We choose
\begin{eqnarray}\label{K0}K_0\ge{\tau_0}^{-1}\log\left(\frac{\mu_0}{\omega}\right)^{-1}
\end{eqnarray}
(where $\tau_0$ corresponds to $\tau$ in the thesis of Theorem~\ref{iteration lemmaOLD}).
 By the thesis of Theorem~\ref{iteration lemmaOLD}, we find a change of coordinates 
 \begin{eqnarray}\label{phi1}
 \phi_1:\ x_1=(\zeta_1, \varphi_1)\in V_{u_1}\to x_0=(\zeta_0, \varphi_0)=\phi_1(\zeta_1, \varphi_1)\in V_{u_0}
 \end{eqnarray}
 where $u_1=\frac{u_0}{2}$,  
 which transforms the vector--field $X_0$ in~\eqref{system0} to
\begin{eqnarray}\label{system1}\label{X1}X_1(x_1)=N_0+ \overline P_0(\zeta_1)+\widetilde
P_1(x_1)\in {\mathcal O}_{u_1}\end{eqnarray}
  $\overline P_0$ is the $\varphi_0$--average of $P_0$
(because in this case $G_+=\Pi_{{\Lambda}_0} T_{K_0} P_{0}=\overline P_0$) and $ \widetilde P_1(x_1)$, corresponding to $P_+$, verifies
\begin{eqnarray*}\VERT \widetilde P_1\VERT _{u_0/2}^{u_0/4}\le {\, \rm const} \frac{\mu_0^2}{\omega}
\end{eqnarray*}
having used~\eqref{K0}. By~\eqref{close to id OLD},~\eqref{C} and~\eqref{gamma0}, the transformation $\phi_1$ in~\eqref{phi1} verifies
\begin{eqnarray}\label{phi1 close to id}
|\phi_1-{\rm id}|_{u_0/2}^{u_0/4}\le \frac{\mu_0}{\omega}\,.
\end{eqnarray}

\paragraph*{Step 2: diagonalization of the linear part} 
The vector--field $X_1(x_1)$ in~\eqref{X1} can be written as
\begin{eqnarray}\label{system1NEW}X_1(x_1)=N_1(\zeta_1)+P_1(x_1)\end{eqnarray}
where  
\begin{eqnarray*}N_1(\zeta_1)=\left(\begin{array}{cc}L\zeta_1\\\omega
\end{array}\right)\,,\quad P_1(x_1):=\breve P(x_1)+\widehat{\overline P}(x_1)+\widetilde P_1(x_1)\end{eqnarray*}
with $\widehat{\overline P}(\zeta_1)$ being the $\varphi_0$--average of $\widehat P$ computed in $\zeta_1$. Here, we have used that $\widetilde P$ has vanishing $\varphi_0$--average and $\breve P$, is $\varphi_0$--independent. 
In Section~\ref{app: Diagonalization} it is shown that the eigenvalues of $L$  are distinct and have negative real part. If $b$ is the $4\times 4$ matrix such that $b^{-1}Lb$ we define the change of coordinates
\begin{eqnarray*}
\phi_2:\quad x_2=(\zeta_2, \varphi_2)\in V_{u_2}\to x_1=(\zeta_1, \varphi_1)=\phi_2(x_2):=(b\zeta_2, \varphi_2)\in V_{u_1}\,.
\end{eqnarray*}
with
\begin{eqnarray*}u_2=(\varepsilon_2, s_2)\,,\quad \varepsilon_2:=\frac{\varepsilon_1}{\|b\|}\,,\qquad s_2:=s_1
\end{eqnarray*}
with $\|b\|$ denoting the operator norm of $b$.
The change $\phi_2$ carries
the vector--field $X_1(x_1)$ in~\eqref{system1NEW} to
 \begin{eqnarray}\label{system2}
X_2(x_2):=\phi_2^{-1}X_1(\phi_2(x_2))=N_2(\zeta_2)+P_2(x_2)
 \end{eqnarray}
where 
 \begin{eqnarray}\label{P(zeta)} N_2(\zeta_2)=\phi_2^{-1}N_1(b\zeta_2)=\left(
\begin{array}{ccc}
\lambda_1 \zeta_{2, 1}\\
\lambda_2 \zeta_{2, 2}\\
\lambda_3 \zeta_{2, 3}\\
\lambda_4\zeta_{2, 4}\\
\omega
\end{array}\right)\,,\qquad P_2(x_2):=\phi_2^{-1}P_1(\phi_2(x_2))\,.\end{eqnarray}
with $\lambda_j$ the eigenvalues of $L$.

 \paragraph*{Step 3: application of Theorem~\ref{normal form lemma}}
Choosing $w_2:=\frac{u_2}{8}$, we have (see~\eqref{system},~\eqref{A} and~\eqref{P(zeta)} )
\begin{eqnarray}\label{P2}\VERT P_2\VERT _{u_2}^{u_2/8}\le {\, \rm const}\max\left\{\frac{\mu_0^2}{\omega}\,,\ \mu_0\frac{r}{a}\,,\ c_1\varepsilon_0^3\,,\ c_2\varepsilon_0^2\right\}=:\mu_1\,.\end{eqnarray}
 We take
 $${{\Lambda}}=\{0\}$$
 so that ${{\Lambda}}_h=\{p_h\}
 $, with $p_h$ as in~\eqref{ph}. We check that the frequency
  $\upomega=(\lambda, {\rm i}\omega)$  is $(\gamma, {{\Lambda}}, 2K_1)$--non resonant with \begin{eqnarray}\label{gamma1}
  |\upomega\cdot k|\ge 
  \gamma_1:=\min\left\{\frac{\upupsilon}{3}\,,\ \omega\right\}\quad \forall\ 0<|k|\in {\mathbb N}
  \end{eqnarray}
 Let $(\alpha_1, \alpha_2, \alpha_3, \alpha_4, k)\in {\mathbb N}^4\times{{\mathbb Z}}\setminus \{(0,0,0,0)\}$.
 If $(\alpha_1, \alpha_2, \alpha_3, \alpha_4)\ne(0,0,0,0)$, then, as ${\, \rm Re\, }\lambda_j<0$ for all $1\le j\le 4$, by~\eqref{bounds lambdai},
 \begin{eqnarray*}
|\alpha\cdot \lambda+{\rm i}\omega k|&=&\left|\sum_{j=1}^4\alpha_j {\, \rm Re\, }\lambda_j+{\rm i}\left(\sum_{j=1}^4\alpha_j\Im\lambda_j+k\omega\right)\right|\ge \left|\sum_{j=1}^4\alpha_j {\, \rm Re\, }\lambda_j\right|=\sum_{j=1}^4\alpha_j|{\, \rm Re\, }\lambda_j|\nonumber\\
&\ge&\min_{1\le j\le 4}|{\, \rm Re\, }\lambda_j|\ge \frac{\upupsilon}{3}\,.
\end{eqnarray*}
If, on the other hand, $(\alpha_1, \alpha_2, \alpha_3, \alpha_4)=(0,0,0,0)$, hence $k\ne 0$, one has $$|\alpha\cdot \lambda+{\rm i}\omega k|=\omega |k|\ge \omega\,.$$
Choosing
\begin{eqnarray*}K_1=\frac{\gamma_1}{2C_*\mu_1\bar\sigma}
\end{eqnarray*}
we have that condition~\eqref{smallcondNEWNEW}  is satisfied and hence Theorem~\ref{normal form lemma} can be applied. As a result, one finds a change of coordinates  \begin{eqnarray}\label{phi3}
\phi_3:\ x_3=(\zeta_3, \varphi_3)\in V_{u_3}\to x_2=(\zeta_2, \varphi_2)=\phi_3(\zeta_3, \varphi_3)\in V_{u_2}
 \end{eqnarray} 
 with $u_3=\frac{u_2}{2}$, which carries the vector--field $X_2(x_2)$ in~\eqref{system2}
to
 \begin{eqnarray}\label{system3}
 X_3(x_3)=N_3(\zeta_3)+P_3(x_3)
 \end{eqnarray}
where
\begin{eqnarray}\label{N3}
N_3(\zeta_3)=N_2(\zeta_3)+G_3(\zeta_3)
\end{eqnarray}
with $N_2$ as in~\eqref{P(zeta)}, and (as $G_3$ satisfies $G_3=\Pi_{\Lambda}T_K G_3$)
\begin{eqnarray}\label{G3}
G_3(\zeta_3)=\left(
\begin{array}{ccc}
\tilde\lambda_1 \zeta_{3, 1}\\
\tilde\lambda_2 \zeta_{3, 2}\\
\tilde\lambda_3 \zeta_{3, 3}\\
\tilde\lambda_4\zeta_{3, 4}\\
\omega
\end{array}\right)
\end{eqnarray}
 Moreover, the following bounds hold:
\begin{eqnarray}\label{G3small}&&\VERT G_3\VERT _{u_3}^{u_3/4}\le 
\VERT \Pi_{{\Lambda}} T_KP_2\VERT _{u_3}^{u_3/4}+\VERT G_3-\Pi_{{\Lambda}} T_KP_2\VERT _{u_3}^{u_3/4}
 \le \mu_1+ \frac{\mu^2_1}{\gamma_1}\le 2\mu_1
\nonumber\\
&&\VERT P_3\VERT _{u_3}^{u_3/4}\le 
\VERT P_2\VERT _{u_2}^{u_2/8}
 e^{-K_1\bar\sigma/4}\le \mu_1
 e^{-\frac{\gamma_1}{8C_*\mu_1}}
 \,.\end{eqnarray}
 By~\eqref{phi* close to id},~\eqref{P2} and~\eqref{gamma1}, the transformation $\phi_3$ in~\eqref{phi3} verifies
 \begin{eqnarray}\label{phi3 close to id}
 |\phi_3-{\rm id}|_{u_2/2}^{u_2/8}\le 2\frac{\mu_1}{\gamma_1}\,.
 \end{eqnarray}
  \paragraph*{Step 4: conclusion} 
By~\eqref{G3small} and Lemma~\ref{Cauchy ineq} below, the numbers $\tilde\lambda_j$  in~\eqref{G3} verify
\begin{eqnarray}\label{tildelambdaj}
    |\tilde\lambda_j|&=&\left|\left.\partial_{\zeta_{3, j}}G_3(\zeta_3)\right|_{\zeta_{3, j}=0}\right|\nonumber\\
    &\le&\frac{\varepsilon_3}{4}\frac{2\mu_1}{\varepsilon_3}=\frac{\mu_1}{2}
\end{eqnarray}
Using~\eqref{system3},~\eqref{N3} and~\eqref{G3},  we have
that the coordinates $\zeta_{3, i}$ satisfy the ODEs
\begin{eqnarray}\label{ODE3}
\dot\zeta_{3, j}=\hat\lambda_j\zeta_{3, j}+P_{3, j}(\zeta_3, \omega t)
\end{eqnarray}
where 
 $\hat\lambda_j:=\lambda_j+\tilde\lambda_j$. 
Moreover, $\hat\lambda_j$ have negative real part, as it follows from~\eqref{tildelambdaj} and the inequality
 \begin{eqnarray*}
   \mu_1\le \gamma_1\le \frac{\upupsilon}{3}\le \min_j|{\, \rm Re\, }\lambda_j|
 \end{eqnarray*}
Rewriting~\eqref{ODE3} in the form
\begin{eqnarray*}
\zeta_{3, j}(t)= \zeta_j(0)e^{\hat\lambda_jt}+\int_0^tP_{3, j}(\zeta_3(\tau), \omega\tau)e^{\hat\lambda_j(t-\tau)}d\tau
\end{eqnarray*}
we find (since ${\, \rm Re\, }\hat\lambda_j<0$)
\begin{eqnarray}\label{asymptotically stable motions}
|\zeta_{3, j}(t)-\zeta_{3, j}(0)e^{\hat\lambda_j t}|&=& 
\left|\int_0^tP_{3, j}(\zeta_3(\tau), \omega\tau)e^{\hat\lambda_j(t-\tau)}d\tau\right|\nonumber\\
&\le& \int_0^{|t|}|P_{3, j}(\zeta_3(\tau), \omega\tau)|d\tau\le \frac{\varepsilon_3}{4}   |t |\mu_1
 e^{-\frac{\gamma_1}{8C_*\mu_1}} \nonumber\\
 &\le& \varepsilon\qquad \forall\ |t|\le \frac{4\varepsilon}{\varepsilon_3\mu_1}\,
 e^{\frac{\gamma_1}{8C_*\mu_1}}\,.
\end{eqnarray}
On the other hand,
taking track of the transformations, $x_0$ and $x_3$ are related via
\begin{eqnarray*}
x_0=\phi_1\circ\phi_2\circ\phi_3(x_3)=\phi_2(x_3)+{\rm O}(\frac{\mu_0}{\omega})+{\rm O}(\frac{\mu_1}{\gamma_1})
\end{eqnarray*}
having used~\eqref{phi1 close to id} and~\eqref{phi3 close to id}.
Taking the projection on $\zeta_3$, we find
\begin{eqnarray*}
\zeta_0(t)=b \zeta_3(t)+{\rm O}(\frac{\mu_0}{\omega})+{\rm O}(\frac{\mu_1}{\gamma_1})
\end{eqnarray*}
Using finally~\eqref{asymptotically stable motions},
we arrive at~\eqref{asymptotically stable}. The theorem is proved with $\varepsilon_*=\varepsilon_3$. $\quad\square$

\section{Proof of Theorems~\ref{iteration lemmaOLD} and~\ref{normal form lemma}}

\subsection{Proof of Theorem~\ref{iteration lemmaOLD}}\label{proof of iteration Lemma}
\begin{definition}
\label{def: time flows}\rm\item[\tiny\textbullet]
  We call {\it time--$\tau$ flow of $Y$} a one--parameter family of
 diffeomorphisms $\Phi_\tau^Y$, $\tau\in {\mathbb R}$,
such that  $x(\tau)=\Phi^Y_\tau(y)$ solves
\begin{eqnarray*}\left\{
\begin{array}{lll}
\partial_{\tau} x=Y(x)\\\\
x(0)=y
\end{array}
\right.\end{eqnarray*}
\item[\tiny\textbullet] For a given $C^{\infty}$ vector--field $Y$, we denote as
\begin{eqnarray*}{\mathcal  L}_Y:=[Y, \cdot]\end{eqnarray*}
 the {\it Lie operator}, where \begin{eqnarray*}[Y, X]:=J_X Y-J_Y X\, ,\quad (J_X)_{ij}:=\partial_{x_j} X_i\end{eqnarray*}
 denotes the {\it Lie brackets}  of two vector--fields. Fixed $\tau>0$, the map \begin{eqnarray}\label{LieOLD}e^{\tau{\mathcal  L}_Y}:=\sum_{{{k}}=0}^{+\infty}\frac{\tau^k}{{{k}}!}{\mathcal  L}^{{k}}_Y\end{eqnarray}is called {\it Lie series} generated by $Y$.

\end{definition}

\begin{proposition}\label{prop: Lie operator}
Assume that 
$e^{\tau{\mathcal  L}_Y}$  is well defined. Then the time--$\tau$ map of $Y$,
 $\Phi^Y_\tau$, carries the ODE~\eqref{oldVF} to $\dot y=Z(y)$, where
$Z=e^{\tau{\mathcal  L}_Y} X$.
\end{proposition}
Proposition~\ref{prop: Lie operator} is a well--known result in differential geometry. A self--contained proof can be however found in Appendix~\ref{Appendix}.

\vskip.1in
\noindent
Our aim is now to provide conditions so that the series~\eqref{LieOLD} is well defined. Without loss of generality, we
 take $\tau=1$. Namely, instead of~\eqref{LieOLD}, we shall use\begin{eqnarray}\label{LieNEW}e^{{\mathcal  L}_Y}:=\sum_{{{k}}=0}^{+\infty}\frac{
{\mathcal  L}^{{k}}_Y
}{{{k}}!}\end{eqnarray}

 \begin{lemma}[Cauchy Inequalities]\label{Cauchy ineq} Let $Z\in{\mathcal O}_{u}$, $u=(\varepsilon, s)$,$0<\rho<\varepsilon$, $0<\sigma<s$.
$${\rm (i)}\ \|\partial^p_{\varphi_i}Z_h\|_{u-\sigma}\le \left(\frac{p}{e \sigma}\right)^p \|Z\|_{u}\,,\qquad{\rm (ii)}\  \|\partial^p_{\zeta_i}Z_h\|_{\varepsilon-\rho, s}\le \frac{p!}{\rho^p} \|Z_h\|_{u}$$
\end{lemma}

\par\medskip\noindent{\bf Proof\ } (i) From the formula
$$\partial^p_{\varphi_i}Z_h=\sum_{(\alpha, k)}z^h_{\alpha, k}\zeta^\alpha ({\rm i}k_i)^p e^{{\rm i}k\cdot \varphi}$$
we get
\begin{eqnarray*}
\|\partial^p_{\varphi_i}Z_h\|_{u-\sigma}&=&\sum_{(\alpha, k)}|z^h_{\alpha, k}|\varepsilon^{|\alpha|_1} |k_i|^p e^{|k|_1(s-\sigma)}\nonumber\\
&\le&\sum_{(\alpha, k)}|z^h_{\alpha, k}|\varepsilon^{|\alpha|_1} |k|_1^p e^{-|k|_1\sigma)}e^{|k|_1s}\nonumber\\
&\le&\frac{1}{\sigma^p}\sup_{x\ge 0}x^p e^{-x}\sum_{(\alpha, k)}|z^h_{\alpha, k}|\varepsilon^{|\alpha|_1} e^{|k|_1s}\nonumber\\
&=&\left(\frac{p}{e \sigma}\right)^p \|Z_h\|_{u},.
\end{eqnarray*}
(ii) From the formula
$$\partial^p_{\zeta_i}Z_h=\sum_{(\alpha, k):\ \alpha_i\ge p}z^h_{\alpha, k}\alpha_i(\alpha_i-1)\cdots(\alpha_i-p+1)\zeta_i^{\alpha_i-p}\prod_{j\ne i}\zeta_j^{\alpha_j}  e^{{\rm i}k\cdot \varphi}$$
we get
\begin{eqnarray*}
\|\partial^p_{\zeta_i}Z_h\|_{\varepsilon-\rho, s}&=&\sum_{(\alpha, k):\ \alpha_i\ge p}|z^h_{\alpha, k}|\alpha_i(\alpha_i-1)\cdots(\alpha_i-p+1)(\varepsilon-\rho)^{\alpha_i-p} (\varepsilon-\rho)^{|\hat\alpha_i|_1} e^{|k|_1s}\nonumber\\
&=&\frac{p!}{\rho^p}\sum_{(\alpha, k):\ \alpha_i\ge p}|z^h_{\alpha, k}|\frac{\alpha_i(\alpha_i-1)\cdots(\alpha_i-p+1)}{p!}(\varepsilon-\rho)^{\alpha_i-p} \rho^p (\varepsilon-\rho)^{|\hat\alpha_i|_1} e^{|k|_1s}
\end{eqnarray*}
with $\hat\alpha_i$ being $\alpha$ deprived if $\alpha_i$. Using now
$$\frac{\alpha_i(\alpha_i-1)\cdots(\alpha_i-p+1)}{p!}(\varepsilon-\rho)^{\alpha_i-p} \rho^p\le \sum_{p=0}^{\alpha_i} \frac{\alpha_i(\alpha_i-1)\cdots(\alpha_i-p+1)}{p!}(\varepsilon-\rho)^{\alpha_i-p} \rho^p=\varepsilon^{\alpha_i}$$
we get the thesis. $\quad \square$
\begin{lemma}\label{Lie brackets}
Let  $w<u\le u_0$;
$Y\in {\mathcal O}_{u_0}$, $W\in {\mathcal O}_{u}$. Then
\begin{eqnarray*}\VERT{\mathcal L}_Y[W]\VERT^{u_0-u+w}_{u-w}\le \VERT Y\VERT^{w}_{u-w}\VERT W\VERT ^{u_0-u+w}_{u}+\VERT W\VERT^{u_0-u+w}_{u-w}\VERT Y\VERT^{u_0-u+w}_{u_0}\, . \end{eqnarray*}
\end{lemma}
\par\medskip\noindent{\bf Proof\ }
One has 
\begin{eqnarray*}
\VERT{\mathcal L}_Y[W]\VERT^{u_0-u+w}_{u-w}&=\VERT J_W Y-J_Y W\VERT^{u_0-u+w}_{u-w}\nonumber\\
&\le  \VERT J_W Y\VERT^{u_0-u+w}_{u-w}+\VERT J_Y W\VERT^{u_0-u+w}_{u-w}
\end{eqnarray*}
Now,  $(J_W Y)_i=\sum_{j}\partial_{x_j} W_i Y_j$, so, using Cauchy inequalities,
\begin{eqnarray*}
 \|(J_W Y)_i\|_{u-w}&\le  \sum_{j}\|\partial_{x_j} W_i\|_{u-w}\| Y_j\|_{u-w}\nonumber\\
 &\le \sum_{j}w_j^{-1}\| W_i\|_{u}\| Y_j\|_{u-w}\nonumber\\
 &=\VERT Y\VERT^w_{u-w}\|W_i\|_{u} 
\end{eqnarray*}
Similarly,
\begin{eqnarray*}\|(J_Y W)_i\|_{u-w}\le \VERT W\VERT_{u-w}^{u_0-u+w}\|Y_i\|_{u_0}\, .  \end{eqnarray*}
Taking the $u_0-u+w$--weighted norms, the thesis follows. $\quad \square$

\begin{lemma}\label{iterateL}
Let $0<w<u$, $Y\in {\mathcal O}_{u+w}$, $W\in {\mathcal O}_{u}$. Then
\begin{eqnarray}\label{q}\VERT{\mathcal L}^{{k}}_Y[W]\VERT^{w}_{u-w}\le {{k}}!q^k \VERT W\VERT^w _{u}\, ,\qquad q:=e \VERT Y\VERT ^w_{u+w} \end{eqnarray}
\end{lemma}

\par\medskip\noindent{\bf Proof\ } We apply Lemma~\ref{Lie brackets} with $W$ replaced by ${\mathcal L}^{i-1}_Y[W]$, 
$u$ replaced by $u-(i-1)w/{{k}}$, 
$w$ replaced by $w/{{k}}$ and, finally, $u_0=u+w$. With $\VERT\cdot\VERT_i^{w}=\VERT\cdot\VERT^{w}_{u-i\frac{w}{{{k}}}}$, $0\le i\le {{k}}$, so that $\VERT\cdot\VERT^w_0=\VERT\cdot\VERT^w_{u}$ and $\VERT\cdot\VERT^w_{{k}}=\VERT\cdot\VERT^w_{u-w}$,
\begin{eqnarray*}
\VERT{\mathcal L}^i_Y[W]\VERT^{w+w/{{k}}}_i&=&\left\VERT\left[Y, {\mathcal L}^{i-1}_Y[W]\right]\right\VERT^{w+w/{{k}}}_i\nonumber\\
&\le&  \VERT Y\VERT^{w/{{k}}}_{i}\VERT{\mathcal L}^{i-1}_Y[W]\VERT^{w+w/{{k}}}_{i-1}+
\VERT Y\VERT^{w+w/{{k}}}_{u+w}\VERT{\mathcal L}^{i-1}_Y[W]\VERT^{w+w/{{k}}}_{i}\,.
\end{eqnarray*}
Hence, de--homogenizating,
\begin{eqnarray*}
\frac{{{k}}}{{{k}}+1}\VERT{\mathcal L}^i_Y[W]\VERT^{w}_i&\le  {{k}} \frac{{{k}}}{{{k}}+1}\VERT Y\VERT^{w}_{i}\VERT{\mathcal L}^{i-1}_Y[W]\VERT^{w}_{i-1}+
\frac{{{k}}^2}{({{k}}+1)^2}\VERT Y\VERT^{w}_{u+w}\VERT{\mathcal L}^{i-1}_Y[W]\VERT^{w}_{i}\nonumber\\
&\le \frac{{{k}}^2}{{{k}}+1}\left(1+\frac{1}{{{k}}+1}\right)\VERT Y\VERT^{w}_{u+w}\VERT{\mathcal L}^{i-1}_Y[W]\VERT^{w}_{i-1}
\end{eqnarray*}
Eliminating the common factor $\frac{{{k}}}{{{k}}+1}$ 
\begin{eqnarray*}
\VERT{\mathcal L}^i_Y[W]\VERT^{w}_i&\le k\left(1+\frac{1}{{{k}}+1}\right)\VERT Y\VERT^{w}_{u+w}\VERT{\mathcal L}^{i-1}_Y[W]\VERT^{w}_{i-1}
\end{eqnarray*}
and 
 iterating ${{k}}$ times from $i={{k}}$ to $i=1$, by Stirling, we get
\begin{eqnarray*}
\VERT {\mathcal L}^{{k}}_Y[W]\VERT^w _{u-w}
&\le   {{k}}^{{k}}\left(1+\frac{1}{{k}}\right)^{{k}}\left(\VERT Y\VERT ^w_{u+w}\right)^{{k}}\VERT W\VERT^w _{u}\le  e^{{k}} {{k}}!\left(\VERT Y\VERT ^w_{u+w}\right)^{{k}}\VERT W\VERT^w _{u}
\end{eqnarray*}
as claimed. $\quad \square$
\vskip.1in
\noindent
Lemma~\ref{iterateL} has the following immediate corollary. We denote as
\begin{eqnarray}\label{tail}e^{{\mathcal L}_Y}_m=\sum_{{{k}}\ge m}\frac{{\mathcal L}^{{k}}_Y}{{{k}}!}\end{eqnarray}
the $m$--tails of the Lie operator~\eqref{LieNEW}.

\begin{proposition}\label{Lie Series}
Let $0<w<u$, $Y\in {\mathcal O}_{u+w}$, $q$ as in~\eqref{q} verify $0\le q<1$.
Then the Lie series $e^{{\mathcal L}_Y}$
defines an operator
\begin{eqnarray*}e^{{\mathcal L}_Y}:\quad  {\mathcal O}_{u}\to {\mathcal O}_{u-w}\end{eqnarray*}
and its $m$--tails~\eqref{tail}
verify
\begin{eqnarray*}\left\VERT e^{{\mathcal L}_Y}_m W\right\VERT^w_{u-w}\le \frac{q^m}{1-q}\VERT W\VERT_{u}^w\qquad \forall\ W\in {\mathcal O}_{u}\, . \end{eqnarray*}
\end{proposition}

\begin{definition}[Homological equation]\rm We call {\it homological equation associated to $N$} an equation of the form \begin{eqnarray}\label{homological equation}[Y, N]=Z\,.\end{eqnarray}
We say that the homological equation is $({\mathcal Z}, {\mathcal Y})$--{\it solvable} if there exist two space of vectorfields ${\mathcal Z}$, ${\mathcal Y}$ such that for any $Z\in{\mathcal Z}$ there exists  $Y\in {\mathcal Y}$ solving~\eqref{homological equation}.
\end{definition}
Recall Definition~\ref{nonresonance}, and, in addition, put the following
\begin{definition}\rm 
Let  ${{\Lambda}}\subset {\mathbb N}^m\times  {\mathbb Z}^n$, with $0\in{{\Lambda}}$.  We say that
$(\lambda, \omega)$ is ${\Lambda}$--{\it resonant} if
$$\alpha\cdot \lambda+{\rm i} k\cdot\omega=0\quad \forall\ (\alpha, k)\in {{\Lambda}}\,.$$
\end{definition}

\begin{proposition}\label{homeq1}
 \item[{\rm (i)}] Let $N\in{\mathcal O}_{u}$ be as in~\eqref{N}, $Y\in{\mathcal O}_{u}$, and assume that the generalized frequencies $(\lambda, \omega)$ are ${\Lambda}$--{\it resonant}. 
Then $Z:={\mathcal L}_YN$ verifies $\Pi_{{\Lambda}}Z=0$, where $\Pi_{{\Lambda}}Z$ is defined as in Definition~\ref{nonresonance}.
 \item[{\rm (ii)}] Let $K\in{\mathbb N}\cup\{\infty\}$; $Z\in {\mathcal O}_u$ be such that  $\Pi_{{\Lambda}}Z=0$, $({\mathbb I}-T_K)Z=0$  and let $(\lambda, {\rm i}\omega)$ be $(\gamma, {{\Lambda}}, K)$--nonresonant. Then there exists a unique $Y\in{\mathcal O}_{u}$
 verifying
 $${\mathcal L}_YN=Z\,,\qquad \Pi_{{\Lambda}}Y=0\,,\ ({\mathbb I}-T_K)Y=0\,.$$
Above, conditions $({\mathbb I}-T_K)Z=0$, $({\mathbb I}-T_K)Y=0$ must be neglected if $K=\infty$.
   \item[{\rm (iii)}] The unique vector--field  $Y$ in {\rm (ii)} verifies
$$\|Y_h\|_{u}\le \frac{\|Z_h\|_{u}}{\gamma}\,.$$

\end{proposition}
\par\medskip\noindent{\bf Proof\ } The Jacobian ${\mathcal D}:=J_N$ of $N$ is given by
$${\mathcal D}=\left(
\begin{array}{cccccc}
D_{m\times m}&0_{m\times n}&\\
0_{n\times m}&0_{n\times n}
\end{array}
\right)\,.$$
Then we have
$$({\cal L}_YN)_h=[Y, N]_h=\Big({\mathcal D} Y-J_Y N(x)\Big)_h=\left[
\begin{array}{lllllll}a_h Y_h-\sum_{j=1}^{m}\lambda_jz_j\partial_{z_j}Y_h-\sum_{i=1}^{n}\omega_i\partial_{\varphi_i}Y_h
\end{array}
\right]_{h}$$
with $a_h=\lambda_h$ if $1\le h\le m$; $a_h=0$ if $m+1\le h\le m+n$.
From these formulae one easily finds the expansion 
$$Z_h=\sum_{\alpha, k}z_{\alpha k}^h \zeta^\alpha e^{{\rm i} k\cdot \varphi}$$
of $Z:={\cal L}_YN$
is given by\begin{eqnarray*}
z_{\alpha k}^{h}={\rm d}_{\alpha k}^{h}y_{\alpha k}^h\end{eqnarray*}
with 
\begin{eqnarray*}
{\rm d}_{\alpha k}^{h}:=\left\{\begin{array}{llll}
-\Big(
\lambda\cdot\alpha+{\rm i}\omega\cdot k-\lambda_h
\Big)
\qquad &{\rm if}\ \ 1\le h\le m\\
-\Big(
\lambda\cdot \alpha+{\rm i}\omega\cdot k
\Big)\qquad &{\rm if}\ \ m+1\le h\le m+n\,.
\end{array}\right.
\end{eqnarray*}
Namely,
\begin{eqnarray*}
{\rm d}_{\alpha k}^{h}=-(\lambda, {\rm i}\omega)\cdot\big((\alpha, k)-p_h\big)
\end{eqnarray*}
where $p_h$ is as in~\eqref{ph}.
As $(\lambda, {\rm i}\omega)$ in $\Lambda$--resonant, 
$z_{\alpha k}^{h}={\rm d}_{\alpha k}^{h}y_{\alpha k}^h=0$ if
$(\alpha, k)-p_h\in{\Lambda}$, namely,
 $\Pi_{{\Lambda}_h}Z_h=0$ for all $1\le h\le n+m$, which amounts to say  $\Pi_{{\Lambda}}Z=0$.  Fix now $Z$ such that $\Pi_{{\Lambda}}Z=0$ and define $Y$ via
$$Y_h=\sum_{\alpha,  k} y_{\alpha k}^h\zeta^\alpha e^{{\rm i}(k\cdot \varphi)}\quad y_{\alpha k}^h:=\frac{z_{\alpha k}^h}{{\rm d}^h_{\alpha k}}$$
As $\Pi_{{\Lambda}}Z=0$, namely, $\Pi_{{\Lambda}_h}Z_h=0$, then $z_{\alpha k}^h=0$ if $(\alpha, k)\in {{\Lambda}_h}$, hence also $y_{\alpha k}^h=0$ if $(\alpha, k)\in {{\Lambda}_h}$, whence $\Pi_{{\Lambda}}Y=0$. 
Similarly, one shows $({\mathbb I}-T_K)Y=0$. If $K<\infty$, then $Y\in \mathcal O_u$ because its Taylor--Fourier series contains only a finite number of terms.
If $K=\infty$, inequality
\begin{eqnarray*}\|Y_h\|_{u}= \sum_{(\alpha,  k)\in{{\Lambda}}_h} \frac{|z_{\alpha k}^h|}{|{\rm d}^h_{\alpha k}|}\varepsilon^{|\alpha|_1} e^{|k|_1 s}\le \frac{\|Z_h\|_{u}}{\gamma}\,,\quad u=(\varepsilon, s)\,.
\end{eqnarray*}
shows that $Y\in \mathcal O_u$. It is obvious that any other $Y'\in \mathcal O_u$ solving  ${\mathcal L}_{Y'}N=Z$ and verifying also $\Pi_{{\Lambda}}Y'=0$ and $({\mathbb I}-T_K)Y=0$
must coincide with $Y$ above.
$\quad\square$

\begin{definition}[Ultraviolet $K$--tail]\rm  Let $K\in{ \mathbb N}$, $K>0$. We say that the vector--field $Z$ is a {\it ultraviolet $K$--tail} if, in the expansion~\eqref{vec expansion}, it is
$$z^h_{\alpha k}=0\quad \forall (\alpha, k)\in{ \mathbb N}^m\times {\mathbb Z}^n:\ |(\alpha, k)|_1<2K\,.$$
\end{definition}
\begin{lemma}[Estimate of the ultraviolet $K$--tail] Let $u=(\varepsilon, s)$, $w=(\rho, \sigma)<u$. Let
$Z\in {\mathcal O}_u$ be a ultraviolet $K$--tail. Then
\begin{eqnarray}\label{tau}\|Z_h\|_{u-w}\le e^{-K\tau}\|Z_h\|_{u}\,,\qquad \tau:=\min\left\{\sigma\,,\ \log(1-\frac{\rho}{\varepsilon})^{-1}\right\}\,.\end{eqnarray}
\end{lemma}
\par\medskip\noindent{\bf Proof\ } By definition,
$$\|Z_h\|_{u-w}=\sum_{|(\alpha, k)|_1\ge 2K} |z_{\alpha k}^h|(\varepsilon-\rho)^{|\alpha|_1} e^{|k|_1 (s-\sigma)}$$
Now, as $|(\alpha, k)|_1=|\alpha|_1+|k|_1$, either $|\alpha|_1\ge K$, or $|k|_1\ge K$.
The terms of the summand with $|\alpha|_1\ge K$
are above by
 $(1-\frac{\rho}{\varepsilon})^K  |z_{\alpha k}^h|\varepsilon^{|\alpha|_1} e^{|k|_1 s}$;
 the ones with
 $|k|_1\ge K$ are bounded by
 $e^{-K\sigma}|z_{\alpha k}^h|\varepsilon^{|\alpha|_1} e^{|k|_1 s}$. $\quad \square$

\begin{lemma}\label{lem: norms}
The norms~\eqref{normXOLD} verify
\begin{eqnarray*}
|X|_u^w\le \VERT X\VERT_u^w\qquad \forall\ X\in {\cal O}_u\,,\ \forall\ 0<w<u\,.
\end{eqnarray*}
\end{lemma}
\par\medskip\noindent{\bf Proof\ } Obvious.
\begin{theorem}\label{iteration lemma}
Let $G$ verify $G=\Pi_{{\Lambda}}T_KG$. The thesis of Theorem~\ref{iteration lemmaOLD} holds also if $X$ in~\eqref{X} is replaced with
\begin{eqnarray*}X=N+G
+P\in {\mathcal O}_{u}\end{eqnarray*}
$G_+$ in~\eqref{G+OLD} with
\begin{eqnarray*}G_+=G+\Pi_{{\Lambda}} T_K P\,.\end{eqnarray*}
and the inequality~\eqref{P+OLD} with
\begin{eqnarray*}
 \VERT P_+\VERT^{w}_{u-2w}
\le \frac{1
}{1-e \gamma^{-1}
\left\VERT P\right\VERT^w_{u}}\left(
e \gamma^{-1}\VERT P\VERT^w_{u}\VERT   P\VERT^w_{u-w}+\left\VERT [Y\,,\ G]\right\VERT^w_{u-w}
+e^{-K \tau}\left\VERT P\right\VERT^w_{u}\right)
\end{eqnarray*}
\end{theorem}
\par\medskip\noindent{\bf Proof\ } 
If
$$P_h=\sum_{(\alpha, k)}p^h_{\alpha, k} \zeta^\alpha e^{\rm i k\cdot\varphi}$$
we let $P_h=P_h^{<2K}+P_h^{\ge 2K}$, with
$$P_h^{<2K}:=\sum_{|\alpha|_1+|k|_1<2K}p^h_{\alpha, k} \zeta^\alpha e^{\rm i k\cdot\varphi}\,,\qquad P_h^{\ge 2K}:=\sum_{|\alpha|_1+|k|_1\ge 2K}p^h_{\alpha, k} \zeta^\alpha e^{\rm i k\cdot\varphi}$$
We have
\begin{eqnarray}\label{X+}
X_+=e^{{\mathcal L}_Y} X=e^{{\mathcal L}_Y}\left(N
+G+P^{<2K}+P^{\ge 2K}\right)=N+G
+P^{<2K}+{\mathcal L}_Y N+P_+
\end{eqnarray}
with
\begin{eqnarray}\label{P+NEW}P_+=e_2^{{\mathcal L}_Y} N
+e_1^{{\mathcal L}_Y}P^{<2K}+e_1^{{\mathcal L}_Y}G+e_0^{{\mathcal L}_Y}P^{\ge 2K}\end{eqnarray}
We further split $P_h^{<2K}=\bar P_h+\tilde P_h^{<2K}$, where \begin{eqnarray*}\bar P_h:=\Pi_{{{\Lambda}_h}}P_h^{<2K}\,,\quad \tilde P_h^{<2K}=\sum_{|\alpha|_1+|k|_1<K\,,(\alpha, k)\notin {{\Lambda}}_h}p^h_{\alpha, k} \zeta^\alpha e^{\rm i k\cdot\varphi}\,.\end{eqnarray*}  Choose $Y\in \tilde{\mathcal O}_{u}$ as the unique solution of \begin{eqnarray}\label{homeq}{\mathcal L}_Y N=-\tilde P^{<2K}\end{eqnarray}
as established by Proposition~\ref{homeq1}.
Then~\eqref{X+} becomes
\begin{align*}X_+=N+G_++P_+
\end{align*}
with $G_+:=G
+\bar P$. The time--one flow of $Y$ is well defined as per Proposition~\ref{Lie Series}, because.

\begin{eqnarray}\label{bound on Y} q:=e \VERT Y\VERT^{w} _{u}&\le  
e \gamma^{-1}\VERT \tilde P^{<2K}\VERT^w_u\le e \gamma^{-1} \VERT P\VERT^w_u<1\, . \end{eqnarray}
By Proposition~\ref{Lie Series}, the Lie series $e^{{\mathcal L}_Y}$
defines an operator
\begin{eqnarray*}e^{{\mathcal L}_Y}:\quad W\in {\mathcal O}_{u-w}\to {\mathcal O}_{u-2w}\end{eqnarray*}
and its tails $e^{{\mathcal L}_Y}_m$
verify
\begin{eqnarray*}
\left\VERT e^{{\mathcal L}_Y}_m W\right\VERT^{w}_{u-2w}&\le  \frac{q^m}{1-q}\VERT W\VERT^{w}_{u-w}\nonumber\\
&\le \frac{\left(e\gamma^{-1}\VERT P\VERT^w_u\right)^m}{1-e\gamma^{-1}\VERT P\VERT^w_u}\VERT W\VERT^{w}_{u-w}
\end{eqnarray*}
for all $W\in {\mathcal O}_{u-w}$.
In particular, $e^{{\mathcal L}_Y}$ is well defined on ${\mathcal O}_{u-w}\subset {\mathcal O}_{u}$, hence $P_+\in {\mathcal O}_{u-w}$. The bounds on $P_+$ in~\eqref{P+NEW} are obtained as follows. Using the homological equation~\eqref{homeq}, one finds
\begin{eqnarray*}
e_2^{{\mathcal L}_Y}N+e_1^{{\mathcal L}_Y}P^{<2K}&=&\sum_{{{k}}=1}^{\infty}\frac{{\mathcal L}^{{{k}}+1}_Y N}{({{k}}+1)!}+
\frac{{\mathcal L}^{{{k}}}_Y P^{<2K}}{{{k}}!}\nonumber\\
&=&\sum_{{{k}}=1}^{\infty}{\mathcal L}^{{{k}}}_Y\left(-\frac{\tilde P^{<2K}}{({{k}}+1)!}+
\frac{ P^{<2K}}{{{k}}!}\right)\nonumber\\
&=&\sum_{{{k}}=1}^{\infty}{\mathcal L}^{{{k}}}_Y\left(\frac{k}{({{k}}+1)!}\tilde P^{<2K}+
\frac{\bar P}{{{k}}!}\right)
\end{eqnarray*}
which gives
\begin{eqnarray*}\label{pertbound1}
\VERT e_2^{{\mathcal L}_Y}N+e_1^{{\mathcal L}_Y}P^{<2K}\VERT_{u-w}^w&\le&
\sum_{{{k}}=1}^{\infty} q^k k!\left\VERT\frac{k}{({{k}}+1)!}\tilde P^{<2K}+
\frac{\bar P}{{{k}}!}\right\VERT_{u-w}^w\nonumber\\
&=&
\sum_{{{k}}=1}^{\infty} q^k k!\left(\frac{k}{({{k}}+1)!}\VERT\tilde P^{<2K}\VERT_{u-w}^w+
\frac{1}{{{k}}!}\VERT{\bar P}\VERT_{u-w}^w\right)\nonumber\\
&\le&
\sum_{{{k}}=1}^{\infty} q^k\VERT P^{<2K}\VERT_{u-w}^w=\frac{q}{1-q}\VERT P^{<2K}\VERT_{u-w}^w
\end{eqnarray*}
The other bounds
\begin{eqnarray*}\label{pertbound2}
&&\left \VERT e_1^{{\mathcal L}_Y}G\right\VERT^{w}_{u-2w}
\le \frac{1}{1-q}\left\VERT {\mathcal L}_Y G\right\VERT^w_{u-w}=\frac{1}{1-q}\left\VERT [Y\,,\ G]\right\VERT^w_{u-w}\nonumber\\
&&
 \left\VERT e_0^{{\mathcal L}_Y}P^{\ge 2K}\right\VERT^{w}_{u-2w}
 \le \frac{1}{1-q}\left\VERT P^{\ge 2K}\right\VERT^w_{u-w}\le\frac{1}{1-q}e^{-K \tau}\left\VERT P\right\VERT^w_{u}
\end{eqnarray*}
are similarly established. 
Finally, it follows from the identity
\begin{eqnarray*}
\phi_+(x_+)=\Phi_1^Y(x_+)=x_++Y(\Phi_{\tau_*}^Y(x_+))\qquad \tau_*\in (0, 1)
\end{eqnarray*}
and Lemma~\ref{lem: norms} that 
\begin{eqnarray*}
|\phi_+-{\rm id}|_{\bar u}^{\bar w}\le  |Y|_{\bar u}^{\bar w}\le  \VERT Y\VERT_{\bar u}^{\bar w}\le  \VERT Y\VERT_{u}^{\bar w} \quad \forall\ \bar u\le u:\ \Phi_{\tau_*}^Y(x_+)\in U_{\bar u}\,,\ \forall \bar w
\end{eqnarray*}
Taking  $\bar u=u-2w$, 
$\bar w=w$ and  using~\eqref{bound on Y}, we have
\begin{eqnarray*}
|\phi_+-{\rm id}|_{u-2w}^w\le  |Y|^w_u\le  \VERT Y\VERT^w_u \le  \gamma^{-1} \VERT P\VERT^w_u
\end{eqnarray*}
which is~\eqref{close to id OLD}. $\quad \square$
\subsection{Proof of  Theorem~\ref{normal form lemma}}\label{proof of NFL}
Put \begin{eqnarray*}x=x_0:=(\zeta_0, \varphi_0)\,,\qquad X_0(x_0):=N(\zeta_0)+P_0(x_0)\,.
\end{eqnarray*} 
We aim to apply Theorem~\ref{iteration lemma} to $X_0$ hence, with $G_0=0$. This is possible because non--resonance condition is verified, and the inequalities~\eqref{Ksigma large} and~\eqref{smallcondNEWNEW} imply~\eqref{smallcond}, provided that $C_*\log(12)\ge e$. We then find $Y_0\in {\mathcal O}_{u}$ such that $\phi_1:=\Phi_1^{Y_0}$ and $\Phi_0=e^{{\mathcal L}_{Y_0}}$ verify
\begin{eqnarray}\label{induction0}
\phi_{1}:\quad x_1\in V_{u-2w}\to x_{0}\in V_{u_{0}}\,,\quad \Phi_0:\ {\mathcal O}_{u}\to {\mathcal O}_{u-2w}\end{eqnarray}
such that
\begin{eqnarray}\label{induction0*}X_1:=e^{{\cal L}_{Y_0}} X_0=N+\bar P_0+P_1\end{eqnarray}
where
\begin{eqnarray}\label{induction0**}\bar P_0\in {\mathcal O}_u\end{eqnarray}
and
\begin{eqnarray}\label{step0}
\VERT P_1\VERT^{w}_{u-2w}
&\le& \frac{1
}{1-e \gamma^{-1}
\left\VERT P_0\right\VERT^w_{u}}\left(
e \gamma^{-1}\VERT P_0\VERT^w_{u}\VERT   P_0\VERT^w_{u-w}+e^{-K \tau}\left\VERT P_0\right\VERT^w_{u}\right)\nonumber\\
&\le&2\left\VERT P_0\right\VERT^w_{u}\left(
e \gamma^{-1}\VERT P_0\VERT^w_{u}+e^{-K \tau}\right)
\end{eqnarray}
If  $\gamma^{-1}\VERT P_0\VERT^w_{u}\le e^{-K \tau}$, there is no much to say. Indeed, using
$$\tau=\min\left\{\sigma\,,\ \log\left(1-\frac{\rho}{\varepsilon}\right)^{-1}\right\}\ge\min\left\{\sigma\,,\ \frac{\rho}{\varepsilon}\right\}= \bar\sigma$$
and~\eqref{Ksigma large}, we have
\begin{eqnarray*}
 \VERT P_1\VERT^{w}_{u-2w}
\le 4e^{-K \tau}\left\VERT P_0\right\VERT^w_{u}=e^{-K \tau+2\log 2}\left\VERT P_0\right\VERT^w_{u}\le
e^{-K\bar\sigma+2\log 2}\left\VERT P_0\right\VERT^w_{u}\le 
 e^{-K \bar\sigma/4}\left\VERT P_0\right\VERT^w_{u}\end{eqnarray*}
 and the proof ends here. 
If, instead, $\gamma^{-1}\VERT P_0\VERT^w_{u}> e^{-K \tau}$, we need a recursion. 	
	\vskip.1in
	\noindent
Fix
\begin{eqnarray}\label{p}p\in{ \mathbb N}\setminus\{0\}\,,\qquad p\le  \frac{K\bar\sigma}{\log(12)}\end{eqnarray}
By~\eqref{Ksigma large}, such a $p$ exist. 
The number $p$ will be used as the amount of iterations. The higher bound in the second inequality in~\eqref{p} will be needed in order to guarantee a suitably fast decay of the perturbing terms. Later on, we shall choose $p$ as the greatest natural number satisfying such inequality, but this is not needed as of now. As of now, we observe that combining  such inequality
with
condition~\eqref{smallcondNEWNEW}, we have
\begin{eqnarray}\label{smallcondNEW}
ep C \gamma^{-1} \VERT P\VERT^{w} _{u}<1\qquad \end{eqnarray}
with $C:=e^{-1}C_*\log(12)$. A suitable  $C\ge 1$ (which corresponds to a suitable $C_*\ge e/\log(12)$) will be fixed along the way.

\vskip.1in
\noindent
{\it Induction} We prove that, if
$$u_0:=u\,,\quad w_0:=w\,,\quad u_j=u-2w-2\frac{j-1}{p}w\,,\quad w_j=\frac{w}{p}\qquad j\in\{1\,,\ldots\,,\ p+1\}$$
for any $j\in\{1\,,\ldots\,,\ p+1\}$, it is possible to find $Y_{j-1}\in {\cal O}_{u_{j-1}}$ such that $\phi_{j}=\Phi^{Y_{j-1}}_1$ and $\Phi_{j-1}:=e^{{\cal L}_{Y_{j-1}}}$
verify
\begin{eqnarray}\label{induction3}
\phi_{j}:\quad x_j\in V_{u_j}\to x_{j-1}\in V_{u_{j-1}}\,,\quad \Phi_{j-1}:\ {\mathcal O}_{u_{j-1}}\to {\mathcal O}_{u_{j}}\end{eqnarray}
and
\begin{eqnarray}\label{induction3*}X_j=\Phi_{j-1}X_{j-1}=N+\sum_{i=0}^{j-1} \bar P_i+P_j\end{eqnarray}
where
\begin{eqnarray}\label{induction4}&&\bar P_{i}\in {\mathcal O}_{u_{i}}\quad \forall\ 0\le i\le j-1\,,\\
\label{induction4*}&& \VERT P_j\VERT^{w}_{u_j}\le \frac{1}{2}\VERT P_{j-1}\VERT^w_{u_{j-1}}\end{eqnarray}
and, moreover,
\begin{eqnarray}\label{induction5}e\gamma^{-1}\VERT P_j\VERT^{w/p}_{u_j}<1\,.\end{eqnarray}
When $j=1$,~\eqref{induction3},~\eqref{induction3*} and~\eqref{induction4} are precisely as in~\eqref{induction0},~\eqref{induction0*} and~\eqref{induction0**}. We
 check that  also~\eqref{induction4*},~\eqref{induction5} are true with $j=1$. Indeed,~\eqref{smallcondNEW} and~\eqref{step0} imply
\begin{eqnarray}\label{induction1}
\VERT P_1\VERT^{w}_{u-2w}
\le 4
e \gamma^{-1}\left(\VERT P_0\VERT^w_{u}\right)^2\le \frac{1}{2}\VERT P_0\VERT^w_{u}\quad (C\ge 8)
\end{eqnarray}
and, moreover,
\begin{eqnarray}\label{induction2}
 e\gamma^{-1}\VERT P_1\VERT^{w/p}_{u-2w}=e\gamma^{-1}\VERT P_1\VERT^{w}_{u-2w}p
\le 
4\left(e \gamma^{-1}\VERT P_0\VERT^w_{u} \right)^2p<\frac{4}{C^2p}<1\end{eqnarray}
so the base step $j=1$ is complete. 
Let us now assume that~\eqref{induction3},~\eqref{induction3*},~\eqref{induction4},~\eqref{induction4*},~\eqref{induction5} hold  for some $j\in \{1\,,\ \ldots\,,\ p\}$, and let us prove the same for $j+1$.		\\
By~\eqref{induction5} and the non--resonance condition, Theorem~\ref{iteration lemma} can be applied with $X=X_j$, $G=\sum_{i=0}^{j-1}\bar P_i$, $P=P_j$, $u=u_j$, $w$ replaced by $ w/p$ and one finds $\Phi_j$ verifying~\eqref{induction3},~\eqref{induction3*},~\eqref{induction4} with $j$ replaced by $j+1$.\\
We  prove that~\eqref{induction4*}  holds with $j$ replaced by $j+1$. This will end the induction, after remarking that 
~\eqref{induction5} with $j$ replaced by $j+1$ is trivially implied by ~\eqref{induction5} itself and~\eqref{induction4*} with $j$ replaced by $j+1$. By the thesis of
 Theorem~\ref{iteration lemma}, we have
\begin{eqnarray*}
\VERT P_{j+1}\VERT^{w/p}_{u_{j+1}}
&\le& \frac{1
}{1-e \gamma^{-1}
\left\VERT P_j\right\VERT^{w/p}_{u_j}}
\nonumber\\
&&
\left(
e \gamma^{-1}\VERT P_j\VERT^{w/p}_{u_j}\VERT   P_j\VERT^{w/p}_{u_j-w/p}+
\VERT[ Y_j\,,\ 
\sum_{i=0}^{j}\bar P_i
]\VERT^{w/p}_{u_j-w/p}
+e^{-K \tau(p)}\VERT P_j\VERT^{w/p}_{u_j}\right)\nonumber\\
&\le&2\left\VERT P_j\right\VERT^{w/p}_{u_j}\left(
e \gamma^{-1}\VERT P_j\VERT^{w/p}_{u_j}+e^{-K \tau(p)}\right)+2\VERT[ Y_j\,,\ 
\sum_{i=0}^{j}\bar P_i
]\VERT^{w/p}_{u_j-w/p}
\end{eqnarray*}
with $\tau(p):=\min\left\{\frac{\sigma}{p}\,,\ \log\left(1-\frac{\rho}{p\varepsilon}\right)^{-1}\right\}$ and 
$\VERT Y_j\VERT^{w/p}_{u_j}\le \gamma^{-1}\VERT P_j\VERT^{w/p}_{u_j}$. We check the following bounds 
\begin{eqnarray}\label{bound1}
&&2 e\gamma^{-1}\VERT P_j\VERT^{w/p}_{u_j}\le  \frac{1}{6}\\
\label{bound2}
&&2 e^{-K\tau(p)}\le \frac{1}{6}\\
\label{bound3}
&&2\VERT[ Y_j\,,\ 
\sum_{i=0}^{j}\bar P_i
]\VERT^{w/p}_{u_j-w/p}\le\frac{1}{6} \VERT P_j\VERT^{w/p}_{u_j}
\end{eqnarray}
 which will imply~\eqref{induction4*} with $j$ replaced by $j+1$ after dehomogeneizating the weight. As a consequence of~\eqref{induction2} (and~\eqref{induction4*} if $j>1$), we have
$$2 e\gamma^{-1}\VERT P_j\VERT^{w/p}_{u_j}\le 2 e\gamma^{-1}\VERT P_1\VERT^{w/p}_{u_1}\le \frac{8}{C^2}\le \frac{1}{6}\qquad C\ge 4\sqrt 3$$
so~\eqref{bound1} is proved.
Moreover, the choice of $p$ in~\eqref{p} guarantees that
$$K\tau(p)=K\min\left\{\frac{\sigma}{p}\,,\ \log\left(1-\frac{\rho}{p\varepsilon}\right)^{-1}\right\}\ge \frac{K\bar\sigma}{p}\ge \log(12)$$
which gives~\eqref{bound2}. It remains to prove~\eqref{bound3}.
Using Lemma~\ref{Lie brackets} with $Y=\bar P_i$, $W=Y_j$, $u_0=u_i$, $u=u_j$, $w$ replaced by $w/p$, we get
\begin{eqnarray*}
2\VERT[ Y_j\,,\ 
\sum_{i=0}^{j}\bar P_i
]\VERT^{w/p}_{u_j-w/p}&\le& 2\sum_{i=0}^{j}\VERT[ Y_j\,,\ 
\bar P_i
]\VERT^{w/p}_{u_j-w/p}\nonumber\\
&\le &2\sum_{i=0}^{j}\VERT P_i\VERT^{w/p}_{u_{j}-w/p}\VERT Y_j\VERT^{2(j-i)w/p+w/p}_{u_j}+\VERT Y_j\VERT^{2(j-i)w/p+w/p}_{u_j-w/p}
\VERT P_i\VERT^{w/p}_{u_{i}}\nonumber\\
&=&2\sum_{i=0}^{j}\frac{1}{2(j-i)+1}\VERT P_i\VERT^{w/p}_{u_{j}-w/p}\VERT Y_j\VERT^{w/p}_{u_j}+\VERT Y_j\VERT^{w/p}_{u_j-w/p}
\VERT P_i\VERT^{w/p}_{u_{i}}\nonumber\\
&\le&4 p\VERT Y_j\VERT^{w/p}_{u_j}\sum_{i=0}^{j}\frac{\VERT P_i\VERT^{w}_{u_i}}{2(j-i)+1}\nonumber\\
&\le&4p\gamma^{-1}\VERT P_j\VERT^{w/p}_{u_j}\sum_{i=0}^{j}\frac{\VERT P_i\VERT^{w}_{u_i}}{2(j-i)+1}\nonumber\\
&=&c\VERT P_j\VERT^{w/p}_{u_j}
\end{eqnarray*}
with
\begin{eqnarray*}
c&:=&4p\gamma^{-1}\sum_{i=0}^{j}\frac{\VERT P_i\VERT^{w}_{u_i}}{2(j-i)+1}=4p\gamma^{-1}\frac{\VERT P_0\VERT^{w}_{u_0}}{2j+1}+8p\gamma^{-1}\VERT P_1\VERT^{w}_{u_1}\le 4p\gamma^{-1}\frac{\VERT P_0\VERT^{w}_{u_0}}{2j+1}+32p
e \gamma^{-2}\left(\VERT P_0\VERT^w_{u}\right)^2\nonumber\\
&\le&\frac{8}{eC}+\frac{32}{peC^2}\le\frac{1}{6}\qquad (C\ge48)
\end{eqnarray*}
This completes the induction. 
Choosing 
now
$$p=p_*:=\left[ \frac{K\bar\sigma}{\log(12)}\right]\qquad  j=p_*+1$$
we obtain
$$X_{*}:=X_{p_*+1}=N+G_*+P_{_*}$$
with $P_*:=P_{p_*+1}$ verifying
$$\VERT P_{*}\VERT^w_{u-4w}\le \frac{1}{2^{p_*+1}}\VERT P_{0}\VERT^w_{u}\le 
2^{-
\frac{K\bar\sigma}{\log(12)}
}\VERT P_{0}\VERT^w_{u}
\le e^{-K\bar\sigma/4}\VERT P_{0}\VERT^w_{u}$$
and $G_*:=\sum_{i=0}^{p_*}\bar P_i$ verifying (by~\eqref{induction1} and~\eqref{induction4*})
$$\VERT G_*-\bar P_0\VERT^w_{u-4w}=\big\VERT \sum_{i=1}^{p_*}\bar P_i\big\VERT^w_{u-4w}\le 2 \VERT P_1\VERT^w_{u-4w}\le  8
e \gamma^{-1}\left(\VERT P_0\VERT^w_{u}\right)^2\,.$$
We finally prove~\eqref{phi* close to id}. By~\eqref{close to id OLD}, the transformations $\phi_j$ in~\eqref{induction3} verify
\begin{eqnarray*}
|\phi_{j}-{\rm id}|_{u_{j-1}-2w_{j-1}}^{w_{j-1}}\le \gamma^{-1}\VERT P_{j-1}\VERT _{u_{j-1}}^{w_{j-1}}\,,\quad j=1\,,\ldots\,,p_*+1
\end{eqnarray*}
Then $\phi_*:=\phi_1\circ\cdots\circ\phi_{p_*+1}$
\begin{eqnarray*}
|\phi_{*}-{\rm id}|_{u-4w}^{w}&\le& \sum_{j=1}^{p_*+1} 
|\phi_{j}-{\rm id}|_{u-4w}^{w}=|\phi_{1}-{\rm id}|_{u-2w}^{w}+
\sum_{j=2}^{p_*+1}  |\phi_{j}-{\rm id}|_{u-4w}^{w}\nonumber\\
&=&|\phi_{1}-{\rm id}|_{u-2w}^{w}+\frac{1}{p_*}
\sum_{j=2}^{p_*+1}  |\phi_{j}-{\rm id}|_{u-4w}^{w_j}\nonumber\\
&\le&
\gamma^{-1}\VERT P_{0}\VERT _{u_{0}}^{w_{0}}+\gamma^{-1}\frac{1}{p_*}
\sum_{j=2}^{p_*+1} \VERT P_{j-1}\VERT _{u_{j-1}}^{w_{j-1}}\nonumber\\
&\le&\gamma^{-1}\VERT P_{0}\VERT _{u_{0}}^{w_{0}}+2\gamma^{-1}\frac{1}{p_*}\VERT P_{1}\VERT _{u_{1}}^{w_{1}}=\gamma^{-1}\VERT P_{0}\VERT _{u_{0}}^{w_{0}}+2\gamma^{-1}\VERT P_{1}\VERT _{u_{0}-2w_0}^{w_{0}}\nonumber\\
&\le&2\gamma^{-1}\VERT P_{0}\VERT _{u_{0}}^{w_{0}}
\end{eqnarray*}
having used~\eqref{induction1} in the last step.
\section{Proof of Proposition~\ref{prop: asymptotically stable}}\label
{app: Diagonalization}
The eigenvalue--eigenvector equation for the matrix $L$, namely,
\begin{eqnarray*}   L y=\lambda y\qquad \lambda\in{ \mathbb C}\,,\ y\in{ \mathbb C}^4\setminus\{0\}
\end{eqnarray*}
can be equivalently formulated as  the request   that the  ODE 
\begin{eqnarray}\label{ODE}
    \dot x(t)=L x(t)
\end{eqnarray}
has the solution  $x(t)=e^{\lambda t}y$.
In turn, writing \begin{eqnarray}\label{xy}
    x=\left(
\begin{array}{lll}
x_1\\
x_1'\\
x_2\\
x_2'
\end{array}
    \right)\,,\quad y=\left(
\begin{array}{lll}
y_1\\
y_1'\\
y_2\\
y_2'
\end{array}
    \right)
\end{eqnarray}
and defining
\begin{eqnarray*}
    {\rm x}:=\left(
\begin{array}{lll}
x_1\\
x_2
\end{array}
    \right)
\end{eqnarray*}
by multiplying the first and the third equation of~\eqref{ODE} by $\epsilon$, $\theta$, respectively, and taking their time--derivative, 
we obtain  the second--order, two--dimensional ODE 
\begin{equation}\label{eq.secsys}
    T\Ddot{{\rm x}}+B\dot{{\rm x}}+V{\rm x}=0,
\end{equation}
where
\begin{eqnarray*}
T:=
\left(\begin{array}{cc}
\epsilon&0\\
0 &\theta
\end{array}\right)\,, \quad 
B:=\theta
\left(\begin{array}{cc}
\epsilon&-\epsilon\\
-\epsilon&\delta
\end{array}\right)\,,\quad V:=2
\left(\begin{array}{cc}
c_1\epsilon&0\\
0&c_2\theta
\end{array}\right)\,.
\end{eqnarray*}
Thus, we equivalently look for  solutions of~\eqref{eq.secsys} of the form
\begin{eqnarray}\label{second order solutions}{\rm x}(t) = e^{\lambda t}{\rm y}\,,\quad {\rm with }\ {\rm y}\in{\mathbb C}^2\setminus\{0\}\end{eqnarray}
up to recover the eigenvector $y$ in~\eqref{xy}
via the relations
\begin{eqnarray*}
   {\rm y}=\left(
\begin{array}{lll}
y_1\\
y_2
\end{array}
    \right)\,,\qquad \left(
\begin{array}{lll}
y'_1\\
y'_2
\end{array}
    \right):=\lambda
    \left(
\begin{array}{lll}
y_1\\
y_2
\end{array}
    \right)\,.
    \end{eqnarray*}
Note that
$T$, $B$ and $V$ are real and symmetric\footnote{
The multiplication by $\epsilon$, $\theta$ allowed to have the matrix  $B$
 symmetric, keeping $T$ and $V$  (diagonal, hence) symmetric.} and their respective  minimum, maximum eigenvalues are given by/satisfy
\begin{eqnarray}\label{eigenvalues bounds}
\lambda^T_-&=&\epsilon\,,\quad \lambda^T_+=\theta\nonumber\\ 
\lambda^B_-&=&\frac{\theta}{2}\left(\epsilon+\delta-\sqrt{(\epsilon+\delta)^2-4\epsilon\upupsilon}\right)\ge\frac{2\theta\epsilon\upupsilon}{\epsilon+\delta} \nonumber\\
 \lambda^B_+&=&\frac{\theta}{2}\left(\epsilon+\delta+\sqrt{(\epsilon+\delta)^2-4\epsilon\upupsilon}\right)\le 
 \theta(\epsilon+\delta)
 \nonumber\\ 
\lambda^V_-&=&2\min\{c_1\epsilon\,,c_2\theta\}\ge2\epsilon\min\{c_1\,,c_2\}\,,\nonumber\\\lambda^V_+&=&2\max\{c_1\epsilon\,,c_2\theta\}\le 2\theta\max\{c_1\,,c_2\}\,.
\end{eqnarray}
  Replacing~\eqref{second order solutions} into~\eqref{eq.secsys} and taking the Hermitian inner product   (here denoted as $(\cdot, \cdot)$)  with ${{\rm y} }$ leads to relation:
\begin{equation*}
    \lambda^2({\rm y}  ,T{\rm y} )+\lambda({\rm y}  ,B{\rm y} )+({\rm y}  ,V{\rm y} )=0.
\end{equation*}
We solve for $\lambda$:
\begin{equation}\label{solution}\lambda = -\frac{({\rm y}  ,B{\rm y} )}{2({\rm y}  ,T{\rm y} )} \pm{\rm i}\frac{\sqrt{
4({\rm y}  ,T{\rm y} )({\rm y}  ,V{\rm y} )-({\rm y}  ,B{\rm y} )^2
}}{2({\rm y}  ,T{\rm y} )}\,.
\end{equation}
As Equation~\eqref{solution} does not change multiplying ${\rm y}$ by an arbitrary  $c\in {\mathbb C}\setminus\{0\}$, we do not loose generality if we assume $({\rm y}  ,{\rm y} )=1$. Under such assumption, by the min-max principle, the expression under the square root is bounded below by
\begin{eqnarray}\label{bound below}
4 \lambda^T_{-} \lambda^{V}_{-}-(\lambda^B_{+})^2\ge8\epsilon^2\min\{c_1\,,c_2\}-\theta^2(\epsilon+\delta)^2>8\epsilon^2\min\{c_1\,,c_2\}-9\theta^2\epsilon^2>0
\end{eqnarray}
 having used~\eqref{friction coefficients},~\eqref{negativity condition} and~\eqref{eigenvalues bounds}.
Equations~\eqref{solution} and~\eqref{bound below} show that the eigenvalues of $L$ come in complex conjugated couples with non--vanishing imaginary part. As we have assumed that the resolvent of the characteristic polynomial of $L$ does not vanish, $L$ has two distinct such couples.  
Moreover, again from~\eqref{friction coefficients},~\eqref{eigenvalues bounds} and~\eqref{solution} , we have
\begin{eqnarray*}
{\, \rm Re\, }\lambda=-\frac{({\rm y}  ,B{\rm y} )}{2({\rm y}  ,T{\rm y} )}\in
\left[-\frac{\lambda_+^B}{2\lambda^T_-}\,,\ -\frac{\lambda^B_-}{2\lambda^T_+}
\right]\subset \left[
-\frac{\theta(\epsilon+\delta)}{2\epsilon}
\,,\ -\frac{\epsilon\upupsilon}{\epsilon+\delta}\right]\subset\left[-\frac{3}{2}\theta\,,-\frac{\upupsilon}{3}\right]
\end{eqnarray*}
which proves~\eqref{bounds lambdai}. $\qquad\square$
\begin{remark}\label{rem: eigenvalues}\rm
The procedure here used to prove Proposition~\ref{prop: asymptotically stable} is considerably simpler than a strategy based on the analysis of the characteristic polynomial of $L$, which is given by $P(\lambda)=(\lambda^2+\theta\lambda+2c_1)(\lambda^2+\delta\lambda+2c_2)-\theta\epsilon\lambda^2$.
Remark that the same argument   may be applied whenever one needs to infer algebraic properties of the eigenvalues of any $n\times n$ matrix $L$ whose ODE~\eqref{ODE} may be put in the form~\eqref{eq.secsys}, with $T$, $B$ and $V$ Hermitian.
\end{remark}

\vskip.1in
\noindent
All the authors contributed equally to this work.\\
The authors declare they do not have conflict of interest.\\
This work has no associated data.

\appendix
\section{Proof of Proposition~\ref{prop: Lie operator}} \label{Appendix}

In general, a diffeomorphism $x=\Phi(y)$ 
transforms the Equation~\eqref{oldVF} to $\dot y=Z(y)$,
where
\begin{eqnarray*}Z(y)=J(y)^{-1}X\big(\Phi (y)\big)\end{eqnarray*}
with $J(y)$ being the Jacobian matrix of the transformation, i.e.,\begin{eqnarray*}J(y)_{hk}=\partial_{y_k}\Phi_h(y)\,,\quad {\rm if}\quad \Phi=(\Phi_1, \ldots, \Phi_n)\,.\end{eqnarray*}
Applying this to $\Phi_\tau^Y$, we obtain that the new vector--field is
\begin{eqnarray*}Z_\tau(y):=J_\tau^Y(y)^{-1}X\big(\Phi^Y_\tau (y)\big)\quad {\rm with} \quad (J_\tau^Y(y))_{hk}:=\partial_{y_k}\big(\Phi^Y_\tau (y)\big)_h \,.\end{eqnarray*}
We stress that the thesis of 
Proposition~\ref{prop: Lie operator}
is an immediate consequence of the following identity
\begin{eqnarray}\label{eq: derivatives}\frac{d^k}{dt^k} Z_t(y)=J_{t}^Y(y)^{-1}{\mathcal  L}^k_Y X\big(\Phi^Y_{t
} (y)\big)\quad \forall\ 0\le t\le \tau\end{eqnarray}
which we are going to prove.
Indeed,~\eqref{eq: derivatives} implies
$$\frac{d^k}{dt^k} Z_t(y)\Big|_{t=0}={\mathcal  L}^k_Y X\big(y\big)$$
which gives
\begin{eqnarray*}Z(y)=Z_\tau(y)=\sum_{k=0}^\infty \frac{\tau^k}{k!}\frac{d^k}{dt^k} Z_t(y)\Big|_{t=0}=\sum_{k=0}^\infty \frac{\tau^k}{k!}{\mathcal  L}^k_Y X\big(y\big)=e^{{\mathcal  L}_Y}X(y)\,.\end{eqnarray*}

\noindent
Let us then prove~\eqref{eq: derivatives}. We use the expansion 
\begin{eqnarray}\label{exp2}\Phi^Y_t  (y)=\Phi^Y_{t _0} (y)+Y\big(\Phi^Y_{t _0} (y)\big)(t -t _0)+o(t -t _0)\end{eqnarray}
and
\begin{eqnarray}\label{inversion}J_t ^Y(y)=\Big({{\mathbb I}}+J_Y\big(\Phi^Y_{t _0} (y)\big)(t -t _0)\Big)J_{t _0}^Y(y)+o(t -t _0)\qquad J_Y(z)_{hk}=\partial_{z_k}Y_h(z)\,.\end{eqnarray}
Equation~\eqref{inversion}
gives
\begin{eqnarray}\label{exp3}(J_t ^Y(\eta))^{-1}=(J_{t _0}^Y(y))^{-1}\Big({{\mathbb I}}-J_Y\big(\Phi^Y_{t _0} (y)\big)(t -t _0)\Big)+o(t -t _0)\,.\end{eqnarray}
While~\eqref{exp2} gives
\begin{eqnarray}\label{expansion}
X\big(\Phi^Y_t  (y)\big)&=&X\big(\Phi^Y_{t _0} (y)+Y\big(\Phi^Y_{t _0} (y)\big)(t -t _0)+o(t -t _0)\big)\nonumber\\
&=&X\big(\Phi^Y_{t _0} (y)\big)+J_X\big(\Phi^Y_{t _0} (y)\big) Y\big(\Phi^Y_{t _0} (y)\big)(t -t _0)+o(t -t _0)
\end{eqnarray}
Collecting~\eqref{exp3} and~\eqref{expansion}, we then find
\begin{eqnarray*}
Z_t (y)&=&J_t ^Y(y)^{-1}X\big(\Phi^Y_t  (y)\big)\nonumber\\
&=&J_{t _0}^Y(y)^{-1}
\nonumber\\
&&
\Big({{\mathbb I}}-J_Y\big(\Phi^Y_{t _0} (y)\big)(t -t _0)+o(t -t _0)\Big)\Big(X\big(\Phi^Y_{t _0} (y)\big)+J_X\big(\Phi^Y_{t _0} (y)\big) Y\big(\Phi^Y_{t _0} (y)\big)(t -t _0)\Big)\nonumber\\
&&+o(t -t _0)\nonumber\\
&=&J_{t _0}^Y(y)^{-1}X\big(\Phi^Y_{t _0} (y)\big)\nonumber\\
&&+J_{t _0}^Y(y)^{-1}\Big(J_X\big(\Phi^Y_{t _0} (y)\big) Y\big(\Phi^Y_{t _0} (y)\big)-J_Y\big(\Phi^Y_{t _0} (y)\big) X\big(\Phi^Y_{t _0} (y)\big)\Big)(t -t _0)+o(t -t _0)
\end{eqnarray*}
This expansion shows that
\begin{eqnarray*}\frac{d}{dt } Z_t (y)
=\frac{d}{dt }\Big(
J_{t }^Y(y)^{-1}X
\big(\Phi^Y_{t } (y)\big)\Big)
=J_{t
}^Y(y)^{-1}{\mathcal  L}_Y X\big(\Phi^Y_{t 
} (y)\big) \end{eqnarray*}
By iteration, we have~\eqref{eq: derivatives}.
$ \qquad \square$

%\small\addcontentsline{toc}{section}{References}
% \bibliographystyle{plain}
%\bibliography{biblio.bib}

\vskip.1in
\noindent

\end{document}